\documentclass[pra,twocolumn,floatfix,superscriptaddress,longbibliography,notitlepage, nofootinbib]{revtex4-1}
\pdfoutput=1

\usepackage[utf8]{inputenc}
\usepackage{braket}
\usepackage{amsmath}
\DeclareMathOperator{\Tr}{Tr}
\usepackage{dcolumn}   
\usepackage{bm}        
\usepackage{amssymb}
\usepackage{mathtools}
\usepackage{tikz}
\usepackage{qcircuit}
\usepackage{appendix}
\usepackage{hyperref}
\usepackage{subcaption}
\usepackage{amsmath,amsfonts,amssymb}
\usepackage{mathtools}
\usepackage{thmtools,thm-restate}
\usepackage{algorithm}
\usepackage{mathrsfs}
\usepackage{pgfplots}
\usetikzlibrary{3d}

\usepackage{algpseudocode}

\newtheorem{definition}{Definition}
\newtheorem{theorem}{Theorem}
\newtheorem{lemma}{Lemma}

\newcommand{\revise}[1]{\textcolor{black}{#1}}
\newcommand{\xbf}{\textbf{x}}

\newcommand{\Ibb}{\mathbb{I}}

\newcommand{\Rbb}{\mathbb{R}}

\usetikzlibrary{arrows.meta}

\def\BibTeX{{\rm B\kern-.05em{\sc i\kern-.025em b}\kern-.08em
    T\kern-.1667em\lower.7ex\hbox{E}\kern-.125emX}}

\begin{document}
\title{Refined Quantum Algorithms for Principal Component Analysis and Solving Linear System}
\author{Nhat A. Nghiem}
\affiliation{Department of Physics and Astronomy, State University of New York at Stony Brook, Stony Brook, NY 11794-3800, USA}
\affiliation{C. N. Yang Institute for Theoretical Physics, State University of New York at Stony Brook, Stony Brook, NY 11794-3840, USA}

\begin{abstract}
  We outline refined versions of two major quantum algorithms for performing principal component analysis and solving linear equations. Our methods are exponentially faster than their classical counterparts and even previous quantum algorithms/dequantization algorithms. Oracle/black-box access to classical data is not required, thus implying great capacity for near-term realization. Several applications and implications of these results are discussed. First, we show that a Hamiltonian $H$ with classically known rows/columns can be efficiently simulated, adding another model in addition to the well-known sparse access and linear combination of unitaries models. Second, we provide a simpler proof of the known result that quantum matrix inversion cannot achieve sublinear complexity $\kappa^{1-\gamma}$ where $\kappa$ is the conditional number of the inverted matrix. 
\end{abstract}
\maketitle

\section{Introduction}
Quantum computation has been undergoing rapid development. Since the early proposals \cite{manin1980computable, benioff1980computer, feynman2018simulating}, tremendous progress has been made in exploring the potential of quantum computers in a wide array of problems. Notable examples include the quantum search algorithm \cite{grover1996fast}, factorization algorithm \cite{shor1999polynomial, regev2023efficient, kitaev1995quantum}, simulation of quantum systems \cite{aharonov2003adiabatic, berry2007efficient,berry2012black,berry2014high,berry2015hamiltonian,berry2015simulating,berry2017quantum, childs2010relationship,childs2018toward, childs2022quantum, lloyd1996universal}, quantum linear equation solving algorithm \cite{harrow2009quantum, childs2017quantum}, and most recently, quantum machine learning/big data algorithm \cite{wiebe2012quantum, lloyd2013quantum,lloyd2016quantum, schuld2014quest,schuld2018supervised,schuld2019evaluating,schuld2019machine,schuld2019quantum,schuld2020circuit, mitarai2018quantum}. 

Among them, quantum principal component analysis (QPCA) \cite{lloyd2014quantum} and quantum linear system solving algorithm (QLSA) \cite{harrow2009quantum,childs2017quantum} stand out as two of the most influential ones, possessing both fundamental and practical impact, because both the PCA and linear system play a vital role in many domains of science and engineering. An efficient quantum solution to these problems, aside from demonstrating a quantum computational advantage, also delivers a meaningful quantum computer's application.  Although both algorithms introduced in \cite{lloyd2014quantum} and \cite{harrow2009quantum} achieve exponential speed-up with respect to their classical counterparts, there exist certain caveats that severely limit their impact. First, a major component inside these algorithms is the so-called oracle/black-box access to the desired classical information. In \cite{lloyd2014quantum,harrow2009quantum}, they assumed to have such an oracle and that this oracle admits efficient implementation. A concrete protocol to realize this oracle was introduced in \cite{giovannetti2008architectures,giovannetti2008quantum}, where the authors proposed the quantum random access memory. However, experimental realization of this architecture has not been achieved in a large-scale, fully functioning form, thus indicating that algorithms with oracle assumption are not suitable to work in the near future, deferring the prospect of quantum advantage. Second, more severely, it was pointed out in the seminal works \cite{tang2018quantum,tang2019quantum,tang2021quantum} that many quantum algorithms, including quantum PCA, only gain speed-up because of the oracle assumption, which turns out to be quite strong. In particular, the authors in \cite{tang2018quantum,tang2019quantum,tang2021quantum} showed that under an analogous assumption, a classical computer can perform PCA with at most a polynomial slowdown compared to \cite{lloyd2014quantum}. More critical discussion can be found in \cite{aaronson2015read}. All in all, a major concern has been raised regarding the potential of quantum computer and particularly its practical utilization.

Recently, it has been shown that quantum computers do not really need strong input assumption to perform PCA and solve linear equations \cite{nghiem2025quantum1}. First, they construct a (new) quantum gradient descent algorithm, based on \cite{nghiem2024simple}. Then they convert the key objective of PCA and linear equation solving to an optimization problem, which can be solved by the quantum gradient descent algorithm. The results are a new QPCA and a new QLSA with logarithmical running time in the dimension, and, in particular, an oracle/black-box assumption is not required. Thus, their results have defied the prevailing belief in the field that strong input assumption is accounted for major quantum speed-up, affirming a positive prospect for quantum advantage.

In this work, on the basis of the above success, we observe that it is possible to improve QPCA and QLSA even exponentially better, in the oracle-free regime. The only information that we require is the classical knowledge of relevant objective, e.g., entries of the featured vectors (in PCA) or entries of the matrix to be inverted in the context of solving linear equations. Our first refined QPCA is based on the power method, which is a simple yet powerful tool to probe the top eigenvalues/eigenvectors of a given matrix. In fact, for principal component analysis, we are also interested in the top eigenvalues/eigenvectors of the so-called covariance matrix, which are referred to as principal components. Therefore, the power method is naturally suited to this objective. As we shall see, by incorporating power method with severally recent advances in quantum algorithms, including \cite{zhang2022quantum} and \cite{gilyen2019quantum}, it is possible to find the principal components with (poly)logarithmic complexity, relative to all parameters, e.g., dimension, number of sample data, and inverse of error tolerance. As a result, our new QPCA achieves a significant improvement over all previous results \cite{lloyd2014quantum, nghiem2025quantum1, nghiem2025new, gordon2022covariance, rodriguez2025quantum, bellante2022quantum, bellante2023quantum}. While the aforementioned approach relying on power method achieves promising performance, we also observe another approach to execute QPCA, which is based on gradient descent algorithm. By reformulating the PCA's objective as an optimization problem, we show that it is possible to execute the gradient descent algorithm using quantum techniques, and thus obtain the top eigenvalues/eigenvectors. As will be discussed later, this approach can be a complement to our previous approach based on the power method. Building on this success, our refined QLSA is a direct extension of our QPCA techniques. The outcome is a new QLSA that achieves (poly)logarithmical dependence on dimension, sparsity, and inverse of error tolerance. Thus, it exhibits an exponential enhancement with respect to the sparsity and inverse of error tolerance compared to existing methods \cite{harrow2009quantum, childs2017quantum, clader2013preconditioned, wossnig2018quantum}. In particular, the inspiration from solving linear equations has guided us to look at another important problem, quantum simulation. The insight is that, Schrodinger's equation -- which describe the dynamics of quantum system -- is essentially a first-order ordinary differential equation and it can be discretized. Therefore, the simulation problem reduces to solving linear equations. By importing the same input information and modifying a step within our QSLA, we show that it is possible to simulate certain Hamiltonian described by a Hermitian matrix of known entries. It thus provides another model for efficient quantum simulation, adding to the well-established sparse access and linear combination of unitaries models \cite{aharonov2003adiabatic,berry2007efficient,berry2012black,berry2014high,berry2015hamiltonian,berry2015simulating,berry2017quantum,low2017optimal,low2019hamiltonian,childs2010relationship,childs2018toward}.

The rest of this work is organized as follows. In Section \ref{sec: overview}, we provide an overview of our main goals and contributions. Specifically, Section \ref{sec: overviewPCA} is devoted to review the problem of principal component analysis, a discussion of existing results and their caveats. We then summarize our new proposal for PCA as a diagram, with a statement of its complexity, compared to the prior results provided in Table \ref{tab: pca}. In Section \ref{sec: overviewlinearsystem}, we do the same thing in the context of solving linear equations, with our results and the existing results summarized in Table \ref{tab: lineareq}. Section \ref{sec: overviewquantumsimulation} contains a background description of the quantum simulation problem and our new model for efficient quantum simulation. A detailed procedure and analysis of our quantum algorithms for PCA, solving linear equations and simulation will be provided in Section \ref{sec: quantumalgorithm} and Section \ref{sec: applicationandimplication}. We remark that our work utilizes many of the results in \cite{gilyen2019quantum}, with important definitions as well as related techniques summarized in the Appendix \ref{sec: prelim}, and we encourage the readers to take a look over these preliminaries before reading the main text. 

\section{Overview of Main Objectives, Prior Results and Our Results}
\label{sec: overview}
In this section, we provide an overview of two key objectives, which are principal component analysis and solving linear algebraic equations. Concurrently, we discuss the progress and results concerning quantum algorithms for these two problems. By pointing out the caveats faced by existing approaches, we accordingly justify the motivation for our main results, which include a new quantum algorithm for performing principal component analysis and solving a system of linear equations that improve state-of-the-art works in many aspects. 

\subsection{Principal Component Analysis}
\label{sec: overviewPCA}
Principal Component Analysis (PCA) is a dimensionality reduction technique widely used in statistics and machine learning. It helps transform high-dimensional data into a lower-dimensional space while preserving as much variance as possible. More formally, let the dataset have $m$ points $\xbf^1,\xbf^2,...,\xbf^m$ where each $\xbf^i \in \Rbb^n$ is a $n$-dimensional vector. Furthermore, for each point $\xbf^i$, we use the subscript $\xbf^i_{j}$ to denote the $j$-th entry, also called the feature of corresponding vector $\xbf^i$. Define the $m \times n$ matrix as:
\begin{align}
    \mathcal{X} = \begin{pmatrix}
        \xbf^1_1 & \xbf^1_2 & \cdots & \xbf^1_n \\
        \xbf^2_1 & \xbf^2_2 & \cdots & \xbf^2_n \\
        \vdots & \vdots & \ddots & \vdots \\
        \xbf^m_1 & \xbf^m_2 & \cdots & \xbf_n^m
    \end{pmatrix}
\end{align}
The centroid of given dataset is defined as $\mu = \sum_{i=1}^m \frac{\xbf^i}{m}$. The centered dataset is formed by subtracting each data point to the mean $\xbf^i \longrightarrow \xbf^i - \mu $. Then upon the subtraction of each data point by $\mu$, we obtain a newly defined matrix $\mathcal{X}_{\rm center} = \mathcal{X} - \mu$. The covariance matrix of given dataset is defined as $  \mathcal{C} = \mathcal{X}_{\rm center}^T\mathcal{X}_{\rm center}$ and as shown in \cite{gordon2022covariance}, it is equivalent to:
\begin{align}
    \mathcal{C}  = \sum_{i=1}^m \frac{1}{m}\xbf^i(\xbf^i)^T - \mu \mu^T = \frac{1}{m} \mathcal{X}^T \mathcal{X} - \mu\mu^T
\end{align}
The essential step of PCA is to diagonalize the above matrix and find the largest eigenvalues with corresponding eigenvectors -- which are called principal components. The projection of given data points $\xbf^1,\xbf^2,...,\xbf^m$ along the top eigenvectors is the newly low-dimensional representation of these points, thus providing a compactification of the given data set. 

Quantum algorithm for principal component analysis was first proposed in \cite{lloyd2014quantum}, and was one of the early influential quantum machine learning algorithms. Their original proposal was designed for simulating density matrix, however, it can be adapted to the context of PCA as well. Following \cite{lloyd2014quantum}, a few variants as well as extensions of quantum PCA have been introduced \cite{gordon2022covariance, rodriguez2025quantum, bellante2022quantum, bellante2023quantum,nghiem2025new, tang2018quantum}. Most recently, in the work \cite{nghiem2025quantum1}, the author showed that quantum computers can perform a wide range of problems without the need for an oracle / black-box, including principal component analysis. To motivate our results, we point out some limitations presented by the aforementioned works. First, in order to apply the main technique in \cite{lloyd2014quantum}, it is required that the covariance matrix, encoded in some density operator, can be efficiently prepared, for example, by means of an oracle / black box. A crucial aspect of PCA is to obtain the centered dataset, and it is not fully justified in \cite{lloyd2014quantum} how to achieve it, in addition to a simple assumption of the already centered dataset provided by the so-called quantum random access memory \cite{giovannetti2008architectures,giovannetti2008quantum, lloyd2013quantum}. Additionally, the method of \cite{lloyd2014quantum} makes use of quantum phase estimation, which resulted in an inevitable complexity being polynomial in the inverse of error. The same assumptions and results are as in~\cite{rodriguez2025quantum, bellante2022quantum, bellante2023quantum}.  The dequantization results by Tang \cite{tang2018quantum,tang2019quantum,tang2021quantum} assume an analogous model to oracle/black-box, so-called \textit{query and access} model that allows efficient $l_2$-norm sampling, and how to realize this model is not known to us. In addition, the dequantization algorithm for PCA, as worked out in \cite{tang2021quantum}, achieves polynomial scaling in the inverse of error.  The recent works of \cite{nghiem2025new, nghiem2025quantum1}, built upon the power method and gradient descent, respectively, do not require an oracle / black box, and their methods achieve a logarithmic dependence on the dimension of the data $n$. However, they have linear scaling in the number of samples $m$ and in the inverse of error. Hence, there is only exponential speed-up (compared to classical algorithm) in dimension $n$. A more technical overview of previous works on quantum PCA is provided in the Appendix \ref{sec: overviewPCA}, therefore, we refer the interested reader to that section for more details.  Subsequently, we will introduce two approaches for performing PCA, one is based on the power method and one is based on gradient descent algorithm. We will see that both these proposals achieve logarithmic scaling in both $n$ and $m$ -- thus yielding an exponential speed-up compared to both classical algorithms and \cite{nghiem2025quantum1}, while do not require oracle/black-box access as in the original work \cite{lloyd2014quantum}. To get a glimpse of how our algorithms work, we provide the following diagram containing the flow and key steps of our two new algorithms for PCA:
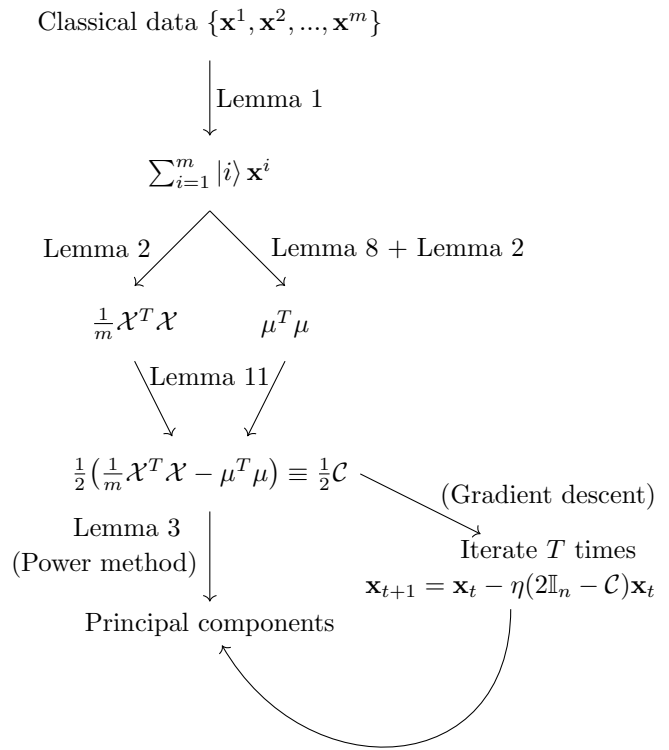
\begin{center}
    \begin{tikzpicture}
        \node at (0,5) {Classical data $\{ \xbf^1,\xbf^2,...,\xbf^m \}$} ;
        \draw[->] (0,4.5) -- (0,3.5) ;
        \node at (0.8,4) { Lemma \ref{lemma: stateprepration}  } ;
        \node at (0,3) { $\sum_{i=1}^m \ket{i} \xbf^i$}; 
        \draw[->] (0,2.5) -- (-1,1.5) ;
        \node at (-1.5, 2) {Lemma \ref{lemma: improveddme}};
        \node at (-1,1) { $ \frac{1}{m}\mathcal{X}^T \mathcal{X}$ } ;
        \node at (2.5, 2) { Lemma \ref{lemma: product} + Lemma \ref{lemma: improveddme} };
        \draw[->] (0,2.5) -- (1,1.5);
        \node at (1,1) { $  \mu^T \mu $}; 

        \node at (0, 0.3) {Lemma \ref{lemma: sumencoding}  };
        \draw[->] (-1, 0.5) -- (-0.5, -0.5) ;
        
        \draw[->] (1,0.5) -- (0.5, -0.5); 
        \node at (0, -1) { $\frac{1}{2}\big( \frac{1}{m}\mathcal{X}^T \mathcal{X}-\mu^T \mu \big) \equiv \frac{1}{2}\mathcal{C} $};

        \draw[->] (0,-1.5) -- (0,-2.7) ;
        \node at (-1.1, -1.7) {Lemma \ref{lemma: largestsmallest}};
        \node at (-1.4, -2.2) {(Power method)}; 

        \draw[->] ( 2, -1) -- (3.6, -1.8) ;
        \node at (4.5, -1.3) {(Gradient descent)};
        \node at (4.5, -2.0) { Iterate $T$ times};
        \node (B) at (4.0,  -2.5) {$\xbf_{t+1} = \xbf_t - \eta (2\Ibb_n -\mathcal{C})\xbf_t$ };
        \node (A) at (0,-3) { Principal components}; 
        \draw[->] (B) to[out = 270, in = 300 ,looseness = 1.5] (A);
    \end{tikzpicture}
\end{center}
To demonstrate the advantage of our proposal, we provide the following table summarizing the relevant complexity in finding the top $2$ eigenvalues/eigenvectors of covariance matrix $\mathcal{C}$ defined above.
\begin{table}[H]
    \centering
    \begin{tabular}{|c|l|}
    \hline
    \textbf{Method} & \textbf{Complexity} \\
    \hline
    First approach (Sec.~\ref{sec: PCApowermethod}) 
    & $ \mathcal{O}\Big( \log(mn) \frac{1}{\Delta^2} \log^2 \big(\frac{n}{\epsilon}\big) \log^2 \frac{1}{\epsilon} \Big)$ \\
    \hline
    Second approach (Sec.~\ref{sec: PCAgradientdescent}) 
    & $ \mathcal{O}\Big( \log^{3} \big(\frac{1}{\epsilon}\big) \big(\frac{4}{\epsilon}\big)^2 \log mn \Big)$ \\
    \hline
    Ref.~\cite{lloyd2014quantum} 
    & $\mathcal{O}\Big(\log (mn) \frac{1}{\epsilon^3}  \Big)$ \\
    \hline
    Ref.~\cite{nghiem2025new} 
    & $\mathcal{O }\Big( m \log(n) \big( \frac{1}{\Delta^2} \log^3 (\frac{n}{\epsilon} ) \frac{1}{\epsilon^2} \big)^2  \Big)$ \\
    \hline
    Ref.~\cite{tang2021quantum} 
    & $\mathcal{O}\Big( \frac{1}{\epsilon^6} + \log(mn) \frac{1}{\epsilon^4} \Big)$ \\
    \hline
    \end{tabular}
    \caption{Table summarizing our result and relevant works of \cite{lloyd2014quantum, nghiem2025new, tang2021quantum}. As we can see, our first approach achieves exponential speed-up with respect to $1/\epsilon$ compared to previous works, meanwhile further exponential speed-up with respect to $m$ (the number of sample data) compared to \cite{nghiem2025new}. }
    \label{tab: pca}
\end{table}

\subsection{System of Linear Algebraic Equations}
\label{sec: overviewlinearsystem}
Linear algebraic equations are vital in many areas of science and engineering. Formally, a $n \times n$ linear system is defined as $A\xbf = \textbf{b}$ with $A$ is some $n\times n$ matrix and $\textbf{b}$ is $n$-dimensional vector. The goal is to find $\xbf$ that satisfies such an equation. Suppose that the system has a unique solution, then it is given by $\xbf = A^{-1} \textbf{b}$. This approach features a direct way, and there is another, indirect approach to solve for the solution of given linear system. By defining a so-called cost function $f(\xbf)= || A\xbf - \textbf{b}||^2$ and seeking its minimum, we can translate the original linear system solving problem into an optimization problem, which can be solved by, for example, gradient descent method. Once we find $\xbf$ such that $f(\xbf)$ is minimized, we can recover the solution to the main linear system $A\xbf = \textbf{b}$. Another more popular choice for the cost function can be $f(\xbf) = \frac{1}{2}||\xbf||^2 +|| A\xbf - \textbf{b}||^2 $, which accounts for the regularization, and it can help speed-up the optimization process.

Quantum algorithm for solving linear algebraic equations was first proposed in \cite{harrow2009quantum}. In quantum context, the definition of ``solving'' linear equations is slightly modified, as quantum algorithms outputs a quantum state $\ket{\xbf} \varpropto A^{-1} \textbf{b}$, i.e., normalization of $A^{-1}\textbf{b}$. In their work, not only they provided a quantum algorithm with exponential speed-up in the dimension (compared to the classical algorithm), they also proved that matrix inversion is BQP-completed, thus ruling out the possibility of efficient classical simulation and dequantization \cite{tang2018quantum,tang2019quantum,tang2021quantum}. Following \cite{harrow2009quantum}, there are many subsequent developments \cite{childs2017quantum, subacsi2019quantum, nghiem2025new2, clader2013preconditioned, huang2019near, zhang2022quantum, wossnig2018quantum} that improve or extend the method in \cite{harrow2009quantum} in different aspects. For example, the work in \cite{childs2017quantum} improves over \cite{harrow2009quantum} in the scaling of the inverse of error, from linear to (poly)logarithmic, thus yielding exponential improvement. The work of \cite{clader2013preconditioned} generalized the method in \cite{harrow2009quantum} to deal with linear system of high conditional number. The Ref.~\cite{wossnig2018quantum} introduced a quantum linear equations solver capable of solving dense system, with quadratic speed-up over \cite{harrow2009quantum}. We remark that an important assumption in most of these works, except \cite{huang2019near} and \cite{zhang2022quantum}, is the oracle/black-box which allows us to query the entries of $A$ efficiently. As we mentioned earlier, this assumption turns out to impose a huge constraint toward the experimental realization of these algorithms, and of quantum advantage in general. The Ref.~\cite{huang2019near} introduced a near-term quantum linear solver, but it is heuristic, thus lacking theoretical performance guarantee. The Ref.~\cite{zhang2022quantum} does not require oracle, but only work when $A$ could be expressed as linear combination of unitaries, $A = \sum_{i=1}^P \alpha_i U_i$ and that each $U_i$ has known implementation mechanism. Most recently, in \cite{nghiem2025quantum1}, the author introduced another approach for solving linear equations based on the gradient descent approach mentioned in the previous paragraph. The method in \cite{nghiem2025quantum1} does not require any kind of oracle access to the entries of $A$, however, their method is only efficient when $A$ is rectangular, i.e., $A$ is a $m \times n$ matrix with $m \ll n$. In subsequent section \ref{sec: solvinglinearequation}, we will provide a new quantum algorithm for solving a linear system, achieving exponential speed-up compared to both classical algorithms and existing quantum linear solvers. For illustration, we provide a diagram showing the main ideas and the flow of the algorithm, with the following definition: $A^i$ refers to the $i$-th row of matrix $A$, and p.s abbreviates positive-semidefinite
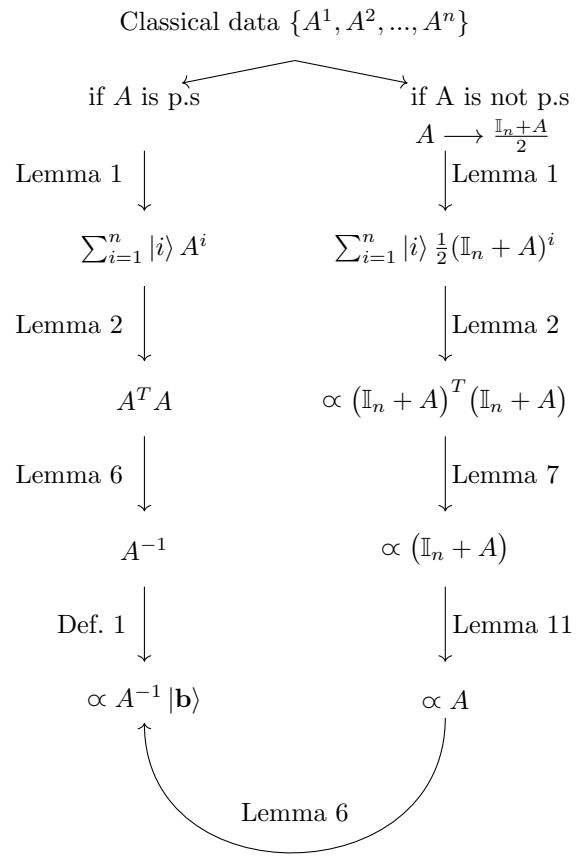
\begin{center}
    \begin{tikzpicture}
         \node at (0,6) {Classical data $\{ A^1,A^2,...,A^n \}$} ;
         \draw[->] (0,5.5) -- (-1.5, 5.2) ;

            \node at (-2.0, 5.0) {if $A$ is p.s};
            \node at (-3.0, 4) {Lemma \ref{lemma: stateprepration}};
            
            \draw[->] (-2,4.3) -- (-2,3.5) ;
        \node at (-2, 3) { $ \sum_{i=1}^n \ket{i} A^i $} ;
        \draw[->] (-2,2.5) -- (-2,1.5);
        \node at (-3.0, 2.0) {Lemma \ref{lemma: improveddme}};
            \node at (-2,1) { $A^T A$};
            \draw[->] (-2,0.5) -- (-2,-0.5) ;
            \node at (-3.0, 0) {Lemma \ref{lemma: negative}};
            \node at (-2,-1) {$A^{-1}$}; 
            \draw[->] (-2,-1.5) -- (-2,-2.5) ;
            \node at (-2.7, -2) {Def.~\ref{def: blockencode}};
            \node (A) at (-2,-3) { $\varpropto A^{-1}\ket{\textbf{b}}$};

        \draw[->] (0,5.5) -- (1.5, 5.2 ) ; 
            \draw[->] (2,4.3) -- (2,3.5); 
            \node at (2.6,5.0) {if A is not p.s };
            \node at ( 2.5 ,4.5) {$A \longrightarrow \frac{\Ibb_n +A}{2}$};
            \node at (2.8,4) {Lemma \ref{lemma: stateprepration}};
        \node at (2,3) { $\sum_{i=1}^n \ket{i} \frac{1}{2}(\Ibb_n+A)^i $  };
        \draw [->] (2,2.5) -- (2,1.5) ;
        \node at (2.8, 2.0) {Lemma \ref{lemma: improveddme}};
        \node at  (2,1) {$ \varpropto \big(  \Ibb_n+A \big)^T \big(  \Ibb_n+A \big)$};
        \draw[->] (2,0.5) -- (2,-0.5) ;
        \node at (2.8, 0.0) {Lemma \ref{lemma: positive}};
        \node at (2,-1) {$\varpropto \big(  \Ibb_n+A \big) $};
        \draw[->] (2, -1.5) -- (2,-2.5) ;
        \node at (2.9, -2.0) {Lemma \ref{lemma: sumencoding}};
        \node (B) at (2,-3) {$ \varpropto A$}; 
        \draw[->] (B) to[out = 270, in = 270 ,looseness = 1.5] (A);
        \node at (0, -4.5) {Lemma \ref{lemma: negative}};
    \end{tikzpicture}
\end{center}
As detailed analysis will be provided later, we provide the following table for comparison of our new proposals versus existing results in the context.
\begin{table}[H]
    \centering
    \resizebox{0.45\textwidth}{!}{
    \begin{tabular}{|c|l|}
    \hline
    \textbf{Method} & \textbf{Complexity} \\
    \hline
    Our method 
    & $\mathcal{O}\Big( \kappa^2 \log(sn) \log^2\big( \frac{\kappa^2}{\epsilon}\big) \log^2 \frac{1}{\epsilon}  \Big)$ \\
    \hline
    Ref.~\cite{nghiem2025new2} 
    & $\mathcal{O}\Big( s^2 \frac{1}{\epsilon}  \big( \log(n) + s^2 \big) \log^{3.5} \frac{s}{\epsilon} \Big)$ \\
    \hline
    Ref.~\cite{harrow2009quantum} 
    & $\mathcal{O}\Big(\frac{1}{\epsilon} s \kappa \log n \Big)$ \\
    \hline
    Ref.~\cite{childs2017quantum} 
    & $\mathcal{O}\Big( s \kappa^2 \log^{2.5} \big(  \frac{\kappa}{\epsilon}\big)  \big(  \log n + \log^{2.5}\frac{\kappa}{\epsilon} \big) \Big)$ \\
    \hline
    Ref.~\cite{clader2013preconditioned} 
    & $\mathcal{O}\Big( \frac{1}{\epsilon}   s^7 \log n \Big)$ \\
    \hline
    \end{tabular}
    }
    \caption{Table summarizing our result and relevant works. Our result achieves exponential improvement with respect to $s$ -- the sparsity of $A$.}
    \label{tab: lineareq}
\end{table}

\subsection{Quantum simulation}
\label{sec: overviewquantumsimulation}
This is probably one of the most promising applications of quantum computers and, in fact, it was one of the key motivations of the quantum computer in the early days \cite{feynman2018simulating}. The dynamic of a quantum system obeys Schrodinger's equation (we set $\hbar = 1$):
\begin{align}
    \frac{\partial \ket{\psi}}{\partial t} = -i H \ket{\psi}
\end{align}
In the time-independent regime, the so-called evolution operator is given by $\exp(-i H t)$, and thus the state of given system at a specific time $t$ is $\ket{\psi}_t = \exp(- i H t) \ket{\psi}_0$ where $\ket{\psi}_0$ is the initial state. The central objective of quantum simulation is to construct, from elementary gates (such as the single qubit or two qubit gate) a unitary $U_t$ such that $\big|\big| U_t - \exp(-i H t) \big|\big| \leq \epsilon$, provided the description of Hamiltonian $H$ of interest. The cost of simulation is typically given by the number of elementary gates used. 

Tremendous progress has been made in this direction \cite{berry2007efficient,berry2012black,berry2014high, aharonov2003adiabatic, childs2010relationship, low2017optimal,low2019hamiltonian, childs2018toward, lloyd1996universal,berry2015hamiltonian,berry2015simulating,tran2021faster, childs2021theory,childs2019nearly,zhao2022hamiltonian}. Two most typical models in the context are, namely, sparse-access model and linear combination of unitaries (LCU) model. In the sparse-access model, the Hamiltonian $H$ is a Hermitian matrix with $s$-sparsity, which means that in each row or column, there are $s$ non-zero entries. In addition, the description of $H$ is provided via two oracles. The first one, upon querying a row index and a number between $1$ and $s$, returns the column index of such non-zero entry. The second one, upon querying a row and column index, returns the value of corresponding entry. As provided in early works \cite{berry2007efficient, aharonov2003adiabatic}, the general strategy to simulate $H$ under this input model, is to decompose $H$ into $H =\sum_j H_j$ where oracle access to each $H_j$ can be obtained from the oracle access to $H$, and the simulation of each $H_j$ can be obtained from single-qubit and two-qubit gates \cite{aharonov2003adiabatic}. The simulation of $H$ is then carried out via product formula, e.g., $\exp(-i H t) \approx \prod_j \exp(-i H_j t)$. In the second, LCU model, Hamiltonian $H$ is given as $H = \sum_{l=1}^L \alpha_l U_l$ -- where $\{ U_l\}$ are unitaries that can be implemented efficiently, and $\{ \alpha \}$ are the coefficients.  According to \cite{berry2015simulating}, one can first divide the time interval $[0,t]$ into, say $r$ segments, and then approximate the evolution operator in each segment as $\exp(-iH \frac{t}{r}) \approx \sum_{k=0}^K  \frac{1}{k!} (-\frac{iHt}{r})^k$. By replacing $H = \sum_{l=1}^L \alpha_l U_l$, further expansion yields $\exp(-iH \frac{t}{r}) \approx \sum_{k=0}^K \sum_{l_1,l_2,...,l_k=1}^L \frac{(-it/r)^k}{k!}  \alpha_{l_1} \alpha_{l_2}.... \alpha_{l_k} U_{l_1} U_{l_2} ... U_{l_k} $. Provided each $U_l$ has a known implementation mechanism, and there is a unitary that generates the state $\varpropto \sum_l \alpha_l\ket{l}$, the Ref.~\cite{berry2015simulating} provides a quantum algorithm to simulate $\exp(-i H \frac{t}{r})$, and by repeating the simulation for $r$ different segments, we can obtain the simulation $\exp(-i H t) = \big(\exp(-i H \frac{t}{r})\big)^r$. 

The complexity of the aforementioned works is logarithmical in the dimension, or in the size of $H$, which highlights the potential of a quantum computer in simulating quantum system, which is expected to be hard for classical devices. Classically, in order to obtain $\exp(-i H t)$, one needs to diagonalize $H$ to find the eigenvalues $\{\lambda_i\}$ and corresponding eigenvectors $\{ \ket{\lambda_i}\} $. The evolution operator can be obtained as $\sum \exp(-i \lambda_i t) \ket{\lambda_i}\bra{\lambda_i}$. Apparently, this approach takes at least linear time, because of the diagonalization step. Furthermore, classically, one needs to know $H$ explicitly in order to perform diagonalization. This fact has inspired us to ask the following question: If we know the entries of $H$ classically, can we perform the quantum simulation ? This input model is different from the two models described above, because neither we are provided with an oracle, nor the Hamiltonian can be expressed as linear combination of unitaries. It turns out that we can efficiently simulate the Hamiltonian provided we know the entries classically. The answer is, in fact, a corollary of the refined quantum linear solving algorithm we outlined in the previous section. Recall that from the diagram, beginning with the classical knowledge of rows of some matrix $A$, at a certain step, we obtain the block encoding of $A$. In this case, if we know the rows of $H$ explicitly, then we can follow the same procedure to construct the block encoding of $H$, from which the simulation $\exp(-i H t)$ can be constructed in a manner similar to \cite{low2017optimal,low2019hamiltonian, gilyen2019quantum}, as we approximate $\exp(-i H t)$ by the Jacobi-Anger polynomial expansion, and use Lemma \ref{lemma: theorem56} to transform $H$ into such a polynomial. This completes a new quantum simulation algorithm with the input model being the classical knowledge of Hamiltonian of interest. For convenience as above, we provide the following diagram, showing the procedure of our quantum simulation algorithm: 
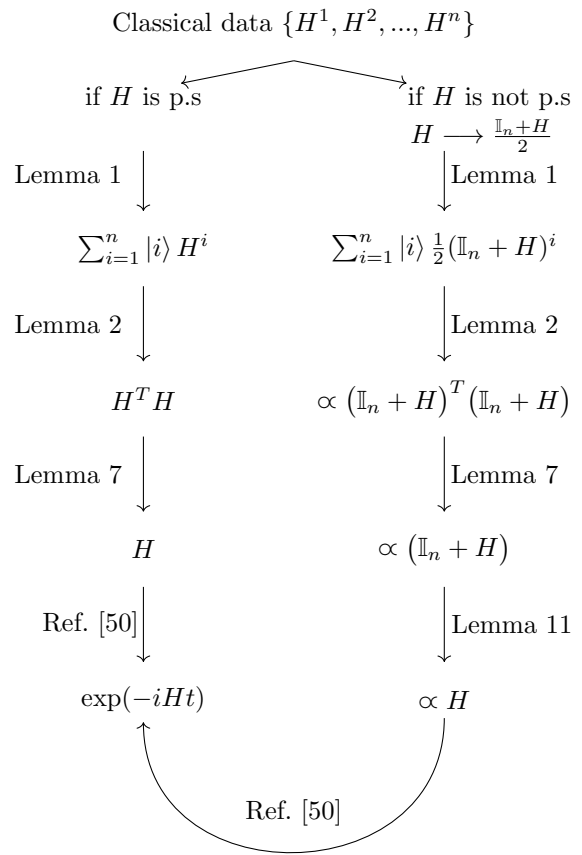
\begin{center}
    \begin{tikzpicture}
         \node at (0,6) {Classical data $\{ H^1,H^2,...,H^n \}$} ;
         \draw[->] (0,5.5) -- (-1.5, 5.2) ;

            \node at (-2.0, 5.0) {if $H$ is p.s};
            \node at (-3.0, 4) {Lemma \ref{lemma: stateprepration}};
            
            \draw[->] (-2,4.3) -- (-2,3.5) ;
        \node at (-2, 3) { $ \sum_{i=1}^n \ket{i} H^i $} ;
        \draw[->] (-2,2.5) -- (-2,1.5);
        \node at (-3.0, 2.0) {Lemma \ref{lemma: improveddme}};
            \node at (-2,1) { $H^T H$};
            \draw[->] (-2,0.5) -- (-2,-0.5) ;
            \node at (-3.0, 0) {Lemma \ref{lemma: positive}};
            \node at (-2,-1) {$H$}; 
            \draw[->] (-2,-1.5) -- (-2,-2.5) ;
            \node at (-2.7, -2) {Ref.~\cite{low2019hamiltonian}};
            \node (A) at (-2,-3) { $\exp(- iH t) $};

        \draw[->] (0,5.5) -- (1.5, 5.2 ) ; 
            \draw[->] (2,4.3) -- (2,3.5); 
            \node at (2.6,5.0) {if $H$ is not p.s };
            \node at ( 2.5 ,4.5) {$H \longrightarrow \frac{\Ibb_n +H}{2}$};
            \node at (2.8,4) {Lemma \ref{lemma: stateprepration}};
        \node at (2,3) { $\sum_{i=1}^n \ket{i} \frac{1}{2}(\Ibb_n+H)^i $  };
        \draw [->] (2,2.5) -- (2,1.5) ;
        \node at (2.8, 2.0) {Lemma \ref{lemma: improveddme}};
        \node at  (2,1) {$ \varpropto \big(  \Ibb_n+H \big)^T \big(  \Ibb_n+ H\big)$};
        \draw[->] (2,0.5) -- (2,-0.5) ;
        \node at (2.8, 0.0) {Lemma \ref{lemma: positive}};
        \node at (2,-1) {$\varpropto \big(  \Ibb_n+ H \big) $};
        \draw[->] (2, -1.5) -- (2,-2.5) ;
        \node at (2.9, -2.0) {Lemma \ref{lemma: sumencoding}};
        \node (B) at (2,-3) {$ \varpropto H$}; 
        \draw[->] (B) to[out = 270, in = 270 ,looseness = 1.5] (A);
        \node at (0, -4.5) {Ref.~\cite{low2019hamiltonian}};
    \end{tikzpicture}
\end{center}

\section{Quantum Algorithm}
\label{sec: quantumalgorithm}
In this section, we first outline a refined quantum PCA algorithm, for which the key technique can be applied to linear solving context in a simple manner. To begin, we recall the following result from \cite{zhang2022quantum}:
\begin{lemma}[Efficient state preparation]
\label{lemma: stateprepration}
    A $n$-dimensional quantum state $\ket{\Phi}$ with known entries (assuming they are normalized to one) can be prepared with a circuit of depth $\mathcal{O}\big( \log (s\log n)\big)$, using $\mathcal{O}(s)$ ancilla qubits ($s$ is the sparsity, or the number of non-zero elements of $\ket{\Phi}$).
\end{lemma}
The above method is probably the most universal amplitude encoding technique, achieving logarithmical depth (in the dimension) meanwhile potentially using linear number of ancillary qubits. In certain cases, a non-sparse state can be prepared using less ancilla qubits, using any of the following methods \cite{grover2000synthesis,grover2002creating,plesch2011quantum, schuld2018supervised, nakaji2022approximate,marin2023quantum,zoufal2019quantum,prakash2014quantum}. By a slight abuse of notation, we define the following $m \times n$ matrix:
\begin{align}
     \mathcal{X} = \begin{pmatrix}
        \xbf^1_1 & \xbf^1_2 & \cdots & \xbf^1_n \\
        \xbf^2_1 & \xbf^2_2 & \cdots & \xbf^2_n \\
        \vdots & \vdots & \ddots & \vdots \\
        \xbf^m_1 & \xbf^m_2 & \cdots & \xbf_n^m
    \end{pmatrix}
\end{align}
Without loss of generalization, we assume that their sum of norms $\sum_{i=1}^m ||\xbf^i ||^2 = 1$, for simplicity. Using the above lemma, suppose that $\mathcal{X}$ is dense (which is typical because each row of $\mathcal{X}$ is the feature vector of data samples, and typically it is dense), we prepare the following state with a circuit $U_\Phi$ of depth $\mathcal{O}(\log mn)$:
\begin{align}
    \ket{\Phi} = \sum_{i=1}^m \sum_{j=1}^n  \xbf^i_j \ket{i} \ket{j}
\end{align}
which is essentially $\sum_{j=1}^n \big( \sum_{i=1}^m \xbf^i_j \ket{i} \big) \ket{j} $. The density state $\ket{\Phi}\bra{\Phi}$ is:
\begin{align}
     \ket{\Phi}\bra{\Phi} = \sum_{j=1}^n  \sum_{k=1}^n \big( \sum_{i=1}^m \xbf^i_j \ket{i}  \big)  \big( \sum_{p=1}^m \xbf^p_k \bra{p}   \big) \otimes  \ket{j} \bra{k}
\end{align}
If we trace out the first register (that holds $\ket{i}$ index), then we obtain the following density state: 
\begin{align}
     \sum_{j=1}^n  \sum_{k=1}^n \big( \sum_{p=1}^m \xbf^p_k \bra{p}   \big)  \big( \sum_{i=1}^m \xbf^i_j \ket{i}  \big) \ket{j}\bra{k}
\end{align}
It is not hard to see that the above matrix is indeed $\mathcal{X}^T \mathcal{X}$ because $\sum_{i=1}^m \xbf^i_j \ket{i}  $ is the $j$-th row of $\mathcal{X}^T$, and $\sum_{p=1}^m \xbf^p_k \bra{p}   $ is the $k$-th column of $\mathcal{X}$. Given that we have $U_\Phi$ that generates $\ket{\Phi}$, by virtue of the following lemma (see their Lemma 45 of \cite{gilyen2019quantum}): 
\begin{lemma}[\cite{gilyen2019quantum} Block Encoding of a Density Matrix]
\label{lemma: improveddme}
Let $\rho = \Tr_A \ket{\Phi}\bra{\Phi}$, where $\rho \in \mathbb{H}_B$, $\ket{\Phi} \in  \mathbb{H}_A \otimes \mathbb{H}_B$. Given unitary $U$ that generates $\ket{\Phi}$ from $\ket{\bf 0}_A \otimes \ket{\bf 0}_B$, then there exists a highly efficient procedure that constructs an exact unitary block encoding of $\rho$ using $U$ and $U^\dagger$ a single time, respectively.
\end{lemma}
it is possible to block-encode $\mathcal{X}^T \mathcal{X}$, with a total circuit of depth $\mathcal{O}( \log mn )$. We can use this block encoding and use Lemma \ref{lemma: scale} with scaling factor $m$, to obtain the block encoding of $ \frac{1}{m}\mathcal{X}^T \mathcal{X}$. \\

The next step is to obtain the block encoding of $\mu \mu^T$. Recall from the above that thanks to Lemma \ref{lemma: stateprepration}, we can prepare the state $ \ket{\Phi}$, which is also $ \sum_{i=1}^m \sum_{j=1}^n  \xbf^i_j \ket{i} \ket{j} = \sum_{i=1}^m \ket{i} \otimes  \xbf^i $. Since $ H^{\otimes \log m} \otimes \Ibb_n $ is simple to prepare, applying it to $\ket{\Phi}$ yields the state $\ket{\Phi'}$, which is:
\begin{align}
   \ket{\Phi'} &=  H^{\otimes \log m} \otimes \Ibb_n  \cdot \sum_{i=1}^m  \ket{i} \otimes \xbf^i \\& = \frac{1}{\sqrt{m}} \ket{0}_{m} \otimes  \big( \sum_{i=1}^m \xbf^i  \big)   + \ket{\rm Redundant}
\end{align} 
where $\ket{0}_m$ specifically denote the first computational basis state of the $m$-dimensional Hilbert space, and $\ket{\rm Redundant}$ is the irrelevant state that is orthogonal to $\ket{0}_{m}\otimes \big( \sum_{i=1}^m \xbf^i  \big)  $. Then Lemma \ref{lemma: improveddme} allows us to construct the block encoding of the density state $\ket{\Phi'}\bra{\Phi'}$, which is equivalent to:
\begin{align}
    \ket{\Phi'}\bra{\Phi'} = \frac{1}{m} \ket{0}_m\bra{0}_m \otimes \big( \sum_{i=1}^m \xbf^i  \big) \big( \sum_{i=1}^m \xbf^i  \big)^T + (...)
\end{align}
where $(...)$ denotes the remaining irrelevant terms. According to Definition \ref{def: blockencode}, the above density operator is again a block encoding of $\frac{1}{m} \big( \sum_{i=1}^m \xbf^i  \big) \big( \sum_{i=1}^m \xbf^i  \big)^T $, which can be combined with Lemma \ref{lemma: scale} (with scaling factor $m$) to transform it into $\frac{1}{m^2} \big( \sum_{i=1}^m \xbf^i  \big) \big( \sum_{i=1}^m \xbf^i  \big)^T  $.  Recall that we have defined the centroid $\mu = \sum_{i=1}^m \frac{\xbf^i}{m} $, so the above procedure allows us to construct the block encoding of $\mu \mu^T$. The complexity of the above procedure is mainly coming from an application of Lemma \ref{lemma: stateprepration} to prepare $\ket{\Phi}$, and of Lemma \ref{lemma: improveddme} to prepare the block encoding of $ \ket{\Phi'}\bra{\Phi'}$, resulting in total complexity  $\mathcal{O}( \log mn)$. 

The block encoding of $\frac{1}{m} \mathcal{X}^T \mathcal{X}$ and of $\mu \mu^T $ allows us to construct the block encoding of $\frac{1}{2}\big( \frac{1}{m} \mathcal{X}^T \mathcal{X} - \mu \mu^T \big)$, which is exactly $\frac{1}{2} \mathcal{C} $ where $\mathcal{C}$ is the covariance matrix. The next goal is to find the principal components -- the largest eigenvalues and the corresponding eigenvectors of $\mathcal{C}$. To this end, we introduce two different approaches, one based on the power method (which was also used in \cite{nghiem2025new}) and one based on gradient descent algorithm.\\
\subsection{Finding principal components based on the power method}
\label{sec: PCApowermethod}
This problem has appeared in a series of works \cite{nghiem2022quantum, nghiem2024improved, nghiem2023improved}, in which they proposed quantum algorithms for finding the largest eigenvalues based on the classical power method \cite{friedman1998error, golub2013matrix}.  In fact, in a recent attempt \cite{nghiem2025new}, the author  also proposed a new quantum PCA algorithm based on power method, however,  as we mentioned previously, their method has linear scaling in the number of sample $m$, and polynomial in the inverse of error.  For the purpose at hand, we refer the interested readers to these original works and recapitulate their main results as follows:
\begin{lemma}
\label{lemma: largestsmallest}
    Given the block encoding of a positive semidefinite Hermitian matrix $A$ of size $n\times n$. Let the eigenvectors of a $A$ be $\ket{A_1}, \ket{A_2},...,\ket{A_n}$ and eigenvalues be $A_1,A_2,...,A_n$. Suppose without loss of generalization that $A_1 > A_2 > ... > A_n $. Then the largest eigenvalue $A_1$ can be estimated up to additive precision $\epsilon$ in complexity $\mathcal{O}\Big(  T_A \big( \frac{1}{ |A_1-A_2| \epsilon}\big) \log \big(\frac{n}{\epsilon} \big) \log \frac{1}{\epsilon} \Big)$ where $T_A$ is the complexity of producing block encoding of $A$. Additionally, the eigenvector $\ket{A_1}$ corresponding to this eigenvalue can be obtained with complexity $\mathcal{O}\Big(  T_A \frac{1}{|A_1-A_2|}\log \big(\frac{n}{\epsilon} \big) \log \frac{1}{\epsilon}\Big) $
\end{lemma}
For the purpose of presentation, we denote the eigenvectors of $\mathcal{C}$ as $\ket{\lambda_1}, \ket{\lambda_2},..., \ket{\lambda_n}$ and the corresponding (ordered) eigenvalues are $\lambda_1 > \lambda_2 > ...> \lambda_n$. The application of the above lemma to our case is straightforward, because the covariance matrix $\mathcal{C}$ is positive semidefinite (see in the previous section, we had $\mathcal{C} = \mathcal{X}_{\rm center}^T\mathcal{X}_{\rm center}$, which is apparently positive semidefinite).  The complexity for obtaining the block encoding of $\frac{1}{2} \mathcal{C}$ is the sum of complexity for obtaining the block encoding of $ \mathcal{X}^T\mathcal{X}$ and of $\mu \mu^T$, so totally it is $\mathcal{O}(\log mn)$. Thus, the complexity in obtaining the first principal component, $\ket{\lambda_1}$, is 
$$\mathcal{O}\Big( \frac{1}{|\lambda_1-\lambda_2|} \log(mn) \log \big(\frac{n}{\epsilon}\big) \log \frac{1}{\epsilon}  \Big)$$
To find the second largest eigenvalue and the corresponding eigenvector, we need the following result, which is an extension of the above Lemma \ref{lemma: largestsmallest}:
\begin{lemma}
\label{lemma: extensionlemmalargestsmallest}
    In the context of Lemma \ref{lemma: largestsmallest}, there is a quantum procedure of complexity $\mathcal{O}\Big( T_A \frac{1}{|A_1-A_2|}\log \big(\frac{n}{\epsilon}\big) \log\frac{1}{\epsilon}\Big) $ that outputs an $\epsilon$-approximated block encoding of $ A_1 \ket{A_1}\bra{A_1}$. 
\end{lemma}
Details of Lemma \ref{lemma: largestsmallest} and the above Lemma \ref{lemma: extensionlemmalargestsmallest} will be provided in the Appendix \ref{sec: reviewpowermethod}. Given that we have the block encoding of $\frac{1}{2}\mathcal{C}$, an application of the above lemma with $A$ replaced by $\mathcal{C}/2$ yields the block encoding of $ \frac{1}{2}\lambda_1 \ket{\lambda_1}\bra{\lambda_1}$. Then we take the block encoding of $\frac{1}{2}\mathcal{C}$, a block encoding of $  \frac{1}{2}\lambda_1\ket{\lambda_1}\bra{\lambda_1}$ and lemma \ref{lemma: sumencoding} to construct the block encoding of:
\begin{align}
    \frac{1}{4}  \Big( \mathcal{C} - \lambda_1 \ket{\lambda_1}\bra{\lambda_1} \Big)  \equiv \mathcal{C}_1
\end{align}
Because the complexity to obtain the block encoding of $\frac{1}{2}\mathcal{C}$ is $\mathcal{O}(\log mn)$, the complexity to obtain the block encoding of $ \frac{1}{2}\lambda_1\ket{\lambda_1}\bra{\lambda_1}$ is $\mathcal{O}\Big( \log(mn) \frac{1}{|\lambda_1-\lambda_2|}  \log \big(\frac{n}{\epsilon}\big) \log\frac{1}{\epsilon} \Big)$. So the complexity to obtain the block encoding of the above operator, $\mathcal{C}_1$, is 
$$ \mathcal{O}\Big( \log(mn) \frac{1}{|\lambda_1-\lambda_2|}  \log \big(\frac{n}{\epsilon}\big) \log\frac{1}{\epsilon} \Big)$$
This matrix $\mathcal{C}_1$ is apparently has the largest eigenvalue to be $\lambda_2/4$ and the corresponding eigenvector is $\ket{\lambda_2}$. So we can repeat an application of Lemma \ref{lemma: largestsmallest} to find them, thus revealing the second principal component. Provided the complexity for block-encoding $\mathcal{C}_1$  as above, the complexity for an application of Lemma \ref{lemma: largestsmallest} is then:
$$ \mathcal{O}\Big( \log(mn) \frac{1}{|\lambda_1-\lambda_2| |\lambda_2-\lambda_3|}  \log^2 \big(\frac{n}{\epsilon}\big) \log^2 \frac{1}{\epsilon} \Big)$$
for obtaining $ \lambda_2/4$ and $\ket{\lambda_2}$.  In a similar manner, we use Lemma \ref{lemma: extensionlemmalargestsmallest} again and repeat the same procedure to obtain the block encoding of $\frac{1}{8} \Big( C - \lambda_1\ket{\lambda_1}\bra{\lambda_1} - \lambda_2 \ket{\lambda_2}\bra{\lambda_2}  \Big) \equiv \mathcal{C}_2$. This operator has $\lambda_3/8$ as largest eigenvalue and corresponding eigenvector is $\ket{\lambda_3}$, which can be found by applying Lemma \ref{lemma: largestsmallest}, resulting in a total complexity:
$$ \mathcal{O}\Big( \log(mn) \frac{1}{|\lambda_1-\lambda_2| |\lambda_2-\lambda_3| |\lambda_3 -\lambda_4  |}  \log^3 \big(\frac{n}{\epsilon}\big) \log^3 \frac{1}{\epsilon} \Big)$$
Continuing the procedure, say, $r$ times to find the $r$ principal components that we desire, then we complete the refined quantum PCA algorithm. We state the main result in the following theorem:
\begin{theorem}
    Given a dataset with $m$ samples and $n$ features 
    \begin{align*}
    \mathcal{X} = \begin{pmatrix}
        \xbf^1_1 & \xbf^1_2 & \cdots & \xbf^1_n \\
        \xbf^2_1 & \xbf^2_2 & \cdots & \xbf^2_n \\
        \vdots & \vdots & \ddots & \vdots \\
        \xbf^m_1 & \xbf^m_2 & \cdots & \xbf_n^m
    \end{pmatrix}
\end{align*}
with the covariance matrix $\mathcal{C}$ as defined above. Let the eigenvectors of $\mathcal{C}$ be $\ket{\lambda_1},\ket{\lambda_2},...,\ket{\lambda_n}$ and corresponding eigenvalues be $\lambda_1 > \lambda_2 > ... > \lambda_n$. Define $\Delta = \max_i \{  |\lambda_i - \lambda_{i+1}| \}_{i=1}^{r}$. The $r$ principal components of $\mathcal{X}$ can be obtained in complexity 
$$ \mathcal{O}\Big( \log(mn) \frac{1}{\Delta^r}   \log^r \big(\frac{n}{\epsilon}\big) \log^r \frac{1}{\epsilon}    \Big)$$
\end{theorem}

In reality, the value of $r$ is typically small, for example $r=2,3$ is common, so our method achieves a polylogarithmic running time on all parameters, providing exponential speed-up compared to previous works \cite{lloyd2014quantum, tang2018quantum,tang2021quantum}.  A crucial factor presented above is $\Delta$, which depends on the gap between eigenvalues. The best regime for this power method-based framework is apparently when $\Delta = \mathcal{O}(1)$. For $\Delta$ being $\frac{1}{\rm polylog(n)}$, our complexity is still efficient. 

The method introduced above achieves polylogarithmic scaling in all parameters, which is major improvement over existing results. However, as we pointed out, the complexity depends on $\Delta$, and if $\Delta$ is polynomially small in the inverse of dimension $n$, the advantage would vanish. In the following section, we introduce another approach, based on redefining the PCA problem as a convex optimization problem, thus can be solved by gradient descent. The complexity of this approach does not depend on the gap $\Delta$, which provides a supplementary framework to this section.  We note that in recent work \cite{nghiem2025quantum1}, the author proposed a new algorithm for PCA, which is also based on gradient descent. However, their approach is technically different from ours, as they encode a vector, say $\xbf = \sum_{i=1}^n x_i \ket{i-1}$ in a diagonal operator $\bigoplus_{i=1}^n x_i$. Here, instead, we embed the vector $\xbf$ into a density matrix-like operator $\xbf \xbf^\dagger$. This strategy has also appeared in recent work \cite{nghiem2025new2}, where the author introduced a new quantum linear solver, also built on gradient descent. In particular, it also appeared in the relevant work \cite{nghiem2023improved}, where they outlined an improved quantum algorithm for gradient descent, aiming at polynomial optimization. In fact, as will be shown below, the function we are going to optimize has the same form as those considered in \cite{nghiem2023improved}, therefore, we can use the same line of reasoning to analyze the complexity. 

\subsection{Finding principal components based on gradient descent}
\label{sec: PCAgradientdescent}
To begin, we remind our readers that we are first interested in the top eigenvector of the covariance matrix $\mathcal{C}$, the eigenvector that corresponds to the largest eigenvalue. Since $\mathcal{C}$ is positive semidefinite and without loss of generalization, we assume that its eigenvalues $\lambda_1,\lambda_2,...,\lambda_n$ are bounded between $0 $ and $1$.  We consider the matrix $\Ibb_n - \mathcal{C}$, which has the eigenvalues $1-\lambda_1,1-\lambda_2,...,1- \lambda_n$ and the same eigenvectors as $\mathcal{C} $.  Given that $1 \geq > \lambda_1 > \lambda_2 > ... > \lambda_n \geq 0$, we have $ 0\leq 1-\lambda_1 < 1- \lambda_2 < ... < 1- \lambda_n \leq 1$, which implies that $\Ibb_n - \mathcal{C}$ is positive-semidefinite and $1-\lambda_1$ is the minimum eigenvalue. To find it, we define $f(\xbf) =  \frac{1}{2}\xbf^T (\Ibb_n-C) \xbf $ and consider the following optimization problem:
\begin{align}
    \min_{\xbf} \ f(\xbf) 
\end{align}
which can be solved by gradient descent algorithm -- a very popular method widely used in many domains of science and engineering. Its execution is simple as the following. First we randomize an initial point $\xbf_0$, then at $t$-th step, iterate the following procedure:
\begin{align}
    \xbf_{t+1} = \xbf_t - \eta \bigtriangledown f(\xbf_t)
\end{align}
where $\eta$ is the hyperparameter. The total iteration step $T$ is typically user-dependent. Some results \cite{nesterov1983method,nesterov2013introductory,boyd2004convex} have established convergence guarantee for the gradient descent algorithm. If the given function is convex, a local minima is also global minima, so by choosing $T = \mathcal{O}\big(  \frac{1}{\epsilon} \big)$ suffices to ensure that $\xbf_T$ is $\epsilon$ close to the true minima of $f(\xbf)$. Meanwhile, for strongly convex functions, $T$ is further improved to $\mathcal{O}\big( \log \frac{1}{\epsilon}\big)$. In our context, the objective function $f(\xbf)$ is convex, as its Hessiasn is $\Ibb_n - \mathcal{C}$ and $\Ibb_n- \mathcal{C}$  is positive-semidefinite. To make things more efficient, we can add a regularization term to $f(\xbf)$, and by a slight abuse of notation, we obtain a new objective function $f(\xbf) = \frac{1}{2} ||\xbf||^2 + \frac{1}{2} \xbf^T (\Ibb_n-C) \xbf$ -- which is a strongly convex function because its Hessian is $\Ibb_n + (\Ibb_n - \mathcal{C})$, which is both lower bounded (by $2-\lambda_1$) and upper bounded (by $2-\lambda_n$).

As mentioned previously, our strategy relies on the embedding of a vector $\xbf$ into a density matrix-like operator, $\xbf \xbf^\dagger$. In this convention, the gradient descent algorithm updates as following:
\begin{align}
    (\xbf \xbf^\dagger)_{t+1} \equiv \xbf_{t+1} \xbf_{t+1}^\dagger
    \end{align}
Given that $\xbf_{t+1} = \xbf_t - \eta \bigtriangledown f(\xbf_t)$ from the regular gradient descent, by a simple algebraic procedure, we have that the above operator is:
\begin{align}
    \xbf_t \xbf_{t+1} -\eta \xbf_t \bigtriangledown^\dagger f(\xbf_t) - \eta \bigtriangledown f(\xbf_t) \xbf_t^\dagger + \eta^2 \bigtriangledown f(\xbf_t) \bigtriangledown^\dagger f(\xbf_t)
\end{align}
Because the function is $f(\xbf) =  \frac{1}{2} ||\xbf||^2 + \frac{1}{2} \xbf^T (\Ibb_n-C) \xbf$, its gradient is:
\begin{align}
    \bigtriangledown f(\xbf)= \xbf +  (\Ibb_n-\mathcal{C}) \xbf  = (2\Ibb_n-\mathcal{C})\xbf
\end{align}
Substituting to the above equation, we obtain:
\begin{align}
      \xbf_{t+1} \xbf_{t+1}^\dagger &= \xbf_t\xbf_t - \eta \xbf_t \xbf_t^\dagger (2\Ibb_n- \mathcal{C})^\dagger- \eta (2\Ibb_n-\mathcal{C}) \xbf_t \xbf_t^\dagger \\ &+ \eta^2 (2\Ibb_n-\mathcal{C}) \xbf_t \xbf_t^\dagger (2\Ibb_n-\mathcal{C})^\dagger 
\end{align}
In the previous section, we have obtained the block encoding of $\frac{1}{2}\mathcal{C}$. As the block encoding of $\Ibb_n$ is simple to prepare (see \ref{def: blockencode}), the block encoding of $\frac{1}{2}\big( \Ibb_n - \frac{1}{2}\mathcal{C}  \big) = \frac{1}{4}\big( 2\Ibb_n - \mathcal{C}\big)$ can be prepared by Lemma \ref{lemma: sumencoding}. Lemma \ref{lemma: scale} can then be used to insert the factor $4\eta$, yielding $\eta \big( 2\Ibb_n - \mathcal{C}\big) $. Suppose that at $t$-th step, we are provided with a block encoding of $\xbf_t \xbf_{t+1}^\dagger $, then we can use Lemma \ref{lemma: product} to construct the block encoding of $\frac{1}{4} \xbf_t \xbf_t^\dagger \big( 2\Ibb_n - \mathcal{C}\big), \frac{1}{4}\big( 2\Ibb_n - \mathcal{C}\big) \xbf_t\xbf_t^\dagger$, and of $\frac{1}{4} \big( 2\Ibb_n - \mathcal{C}\big) \xbf_t \xbf_t^\dagger \big( 2\Ibb_n - \mathcal{C}\big)$. Then we use \ref{lemma: sumencoding} to construct the block encoding of:
\begin{align}
    \frac{1}{4} \Big( \xbf_t\xbf_t - \eta \xbf_t \xbf_t^\dagger (2\Ibb_n- \mathcal{C})^\dagger- \eta (2\Ibb_n-\mathcal{C}) \xbf_t \xbf_t^\dagger \\ + \eta^2 (2\Ibb_n-\mathcal{C}) \xbf_t \xbf_t^\dagger (2\Ibb_n-\mathcal{C})^\dagger  \Big)
    \label{17}
\end{align}
which is exactly $\frac{1}{4} \xbf_{t+1}\xbf_{t+1}^\dagger$, and the factor $4$ can be removed using Lemma \ref{lemma: amp_amp}. Thus, beginning with some initial operator $\xbf_0 \xbf_0^\dagger$, which can be block-encoded by simply using Lemma \ref{lemma: improveddme} with an arbitrary unitary $U_0$ that generates the state $\ket{\bf 0} \xbf_0 + \ket{\rm Redundant}$, then we can iterate the above procedure for a total of $T$ times, which produces the block encoding of  $\xbf_T \xbf_T^\dagger$. To obtain the state $\ket{\xbf_T}$, we take such block encoding and apply it to some state $\ket{\alpha}$, according to definition \ref{def: blockencode}, we obtain the state $ \ket{\bf 0} (\xbf_T \xbf_T^\dagger) \ket{\alpha} + \ket{\rm Garbage}$. Measurement of the ancilla and post-select in $\ket{\bf 0}$ yields the state $\ket{\xbf_T}$.

To analyze the complexity, we recall that the complexity for producing the (scaled) covariance matrix $\mathcal{C}/2$ is $\mathcal{O}( \log mn)$.  Thus, the complexity in obtaining the block encoding of $\eta (2\Ibb_n - \mathcal{C})$ is the same, $\mathcal{O}( \log mn)$. Let $\mathcal{T}_t$ denote the complexity of obtaining the block encoding of $\xbf_t\xbf_t^\dagger$. In Eqn.~\ref{17}, the operator $\xbf_t\xbf_t^\dagger$ appears 4 times, the operator  $\varpropto  (2\Ibb_n - \mathcal{C})$ appears 4 times, therefore, the complexity for producing block encoding of $ \frac{1}{4}\xbf_{t+1} \xbf_{t+1}^\dagger$ is $4 \mathcal{O}(\log mn ) + 4 \mathcal{T}_t$. An application of Lemma \ref{lemma: amp_amp} to transform $\frac{1}{4}\xbf_{t+1} \xbf_{t+1}^\dagger \longrightarrow \xbf_{t+1} \xbf_{t+1}^\dagger $ incurs further complexity $\mathcal{O}( \log \frac{1}{\epsilon}\big)$. So the complexity for producing block encoding of $ \xbf_{t+1} \xbf_{t+1}^\dagger$ is 
$$\mathcal{T}_{t+1} = 4 \mathcal{O}\big(  \log (mn) \log (\frac{1}{\epsilon}) \big)  + 4 \log (\frac{1}{\epsilon}) \mathcal{T}_t  $$ Using induction, we have 
$$\mathcal{T}_t =4 \mathcal{O}\big( \log (\frac{1}{\epsilon})\log (mn) \big) + 4 \log (\frac{1}{\epsilon}) \mathcal{T}_{t-1}  $$ and thus 
\begin{align}
    \mathcal{T}_{t+1} &=  \big( 4\log (\frac{1}{\epsilon})  + 4^2 \log^2 (\frac{1}{\epsilon})\big) \mathcal{O}( \log mn) +  4^2 \log^2 (\frac{1}{\epsilon}\big) \mathcal{T}_{t-1}
\end{align}
Continuing the process, we have 
\begin{align}
    \mathcal{T}_{t+1} &= \Big( \big( 4\log \frac{1}{\epsilon} \big) + \big( 4\log \frac{1}{\epsilon} \big) ^2 + ... +\big( 4\log \frac{1}{\epsilon} \big) ^{t+1} \Big) \\ & \times  \mathcal{O}(\log mn) +\big( 4\log \frac{1}{\epsilon} \big) ^{t+1} \mathcal{T}_0
\end{align} 
where $\mathcal{T}_0$ is the complexity for producing $\xbf_0\xbf_0^\dagger$ , which is $\mathcal{O}(\log n)$ due to an application of Lemma \ref{lemma: improveddme}. So for a total of $T$ iteration steps, the complexity is $\mathcal{O}\big( \big( 4\log \frac{1}{\epsilon} \big) ^T \log mn \big)$.  Because our objective function $f(\xbf)$ is strongly convex as pointed out before, the value of $T$ can be $\mathcal{O}\big( \log \frac{1}{\epsilon}\big) $, yielding a final complexity $ \mathcal{O}\Big(  4\log \big(\frac{1}{\epsilon} \big)  \frac{1}{\epsilon} \log mn  \Big) $ for producing $\ket{\xbf_T}$, which is an approximation to $\ket{\lambda_1}$ -- the eigenvector corresponding to the largest eigenvalue of $\mathcal{C}$. 

Now we show how to find the next eigenvector, $\ket{\lambda_2}$. We use the same strategy as in the previous section, where our aim was to find the top eigenvector of $\mathcal{C}- \lambda_1 \ket{\lambda_1}\bra{\lambda_1}$, so we need a tool similar to Lemma \ref{lemma: extensionlemmalargestsmallest} . In the Appendix \ref{sec: reviewpowermethod}, \ref{sec: extensiongradientdescent} we show the following:
\begin{lemma}\label{lemma: extensiongradientdescent}
\
    Given that the block encoding of $\xbf_T \xbf_T^\dagger $ can be obtained by the above procedure, there is a quantum procedure that outputs an $\epsilon$-approximated block encoding of $\lambda_1 \ket{\lambda_1}\bra{\lambda_1}$. The complexity of this procedure  is $\mathcal{O}\big(4 \log^2(\frac{1}{\epsilon}) \frac{1}{\epsilon} \cdot \log mn  \big) $
\end{lemma}
The (approximated) block-encoded operator $\lambda_1 \ket{\lambda_1}\bra{\lambda_1} $ can be transformed into $ \frac{1}{2 }\lambda_1 \ket{\lambda_1}\bra{\lambda_1}  $ simply using Lemma \ref{lemma: scale}, which can then be used with the already-have block encoding of $\frac{1}{2}\mathcal{C}$ and Lemma \ref{lemma: sumencoding}   to construct the block encoding of $\varpropto \big(  \mathcal{C} - \lambda_1 \ket{\lambda_1}\bra{\lambda_1} \big)$. As discussed in previous section, this operator has $\lambda_2$ being the maximum eigenvalue and corresponding eigenvector is $\ket{\lambda_2}$, so we can first convert it to a convex optimization problem as we did from the beginning of Section \ref{sec: PCAgradientdescent}, and then repeat the  procedure as above, to find $\ket{\lambda_2}$, $\ket{\lambda_3},..., \ket{\lambda_r}$ -- the $r$ principal components. We summary the result of this section in the following theorem.
\begin{theorem}
    Given a dataset with $m$ samples and $n$ features 
    \begin{align*}
    \mathcal{X} = \begin{pmatrix}
        \xbf^1_1 & \xbf^1_2 & \cdots & \xbf^1_n \\
        \xbf^2_1 & \xbf^2_2 & \cdots & \xbf^2_n \\
        \vdots & \vdots & \ddots & \vdots \\
        \xbf^m_1 & \xbf^m_2 & \cdots & \xbf_n^m
    \end{pmatrix}
\end{align*}
with the covariance matrix $\mathcal{C}$ as defined above. Let the eigenvectors of $\mathcal{C}$ be $\ket{\lambda_1},\ket{\lambda_2},...,\ket{\lambda_n}$ and corresponding eigenvalues be $\lambda_1 > \lambda_2 > ... > \lambda_n$. The $r$ principal components of $\mathcal{X}$ can be obtained in complexity 
$$ \mathcal{O}\Big(  \log^{2r-1} (\frac{1}{\epsilon}) \big(\frac{4}{\epsilon}\big)^r \log mn \Big) $$
\end{theorem}
Comparing to the complexity of the previous section, we can see that this gradient descent-based approach does not depend on the gap $\Delta$ between eigenvalues, as we expected. However, there is a trade-off on the inverse of error, as this approach exhibits polynomial dependence on $\frac{1}{\epsilon}$. 
\subsection{Solving linear algebraic equations}
\label{sec: solvinglinearequation}
The linear system is defined as $A\xbf = \textbf{b}$. Similarly to previous contexts \cite{harrow2009quantum, childs2017quantum}, we assume without loss of generality that $A$ is $s$-sparse Hermitian and its eigenvalues are falling between $(-1,1)$. Suppose that a unique solution exists, it is given by $\xbf = A^{-1} \textbf{b}$. For concreteness, we further define:
\begin{align}
    A = \begin{pmatrix}
        A_{11} & A_{12} &  \cdots & A_{1n} \\
        A_{21} & A_{22} & \cdots & A_{2n} \\
        \vdots & \vdots & \ddots & \vdots \\
        A_{n1} &  A_{n2} & \cdots & A_{nn}
    \end{pmatrix}, \ \textbf{b }= \begin{pmatrix}
        b_1 \\
        b_2\\
        \vdots \\
        b_n
    \end{pmatrix}
\end{align}
In quantum context, the goal is to obtain the state $\ket{\xbf} \varpropto A^{-1} \textbf{b}$. We recall that at the beginning of Section.~\ref{sec: quantumalgorithm}, we showed how to construct the block encoding of $\mathcal{X}^T \mathcal{X}$ (see the discussion above Lemma \ref{lemma: improveddme}), and the same technique can be used to construct the block encoding of $A^T A$. More specifically, we first use Lemma \ref{lemma: stateprepration} create the state:
\begin{align}
    \ket{\Phi} = \sum_{i=1}^n \sum_{j=1}^n A_{ij }\ket{i}\ket{j} 
\end{align}
in complexity $\mathcal{O}(\log sn) = \mathcal{O}(\log sn)$. The reason for the appearance of $s$ -- the sparsity of $A$, is because by definition, it is the maximum number of non-zero entries in each row or column of $A$. By tracing out the first register that holds $\{ \ket{i}\} $ of the above state, we obtain the density state $A^T A$, which can be block-encoded via Lemma \ref{lemma: improveddme}. To proceed, we point out the following result from \cite{gilyen2019quantum}:
\begin{lemma}[Negative Power Exponent \cite{gilyen2019quantum}, \cite{chakraborty2018power}]
\label{lemma: negative}
    Given a block encoding of a positive matrix $\mathcal{M}$ such that 
    $$ \frac{\Ibb}{\kappa_M} \leq \mathcal{M} \leq \Ibb. $$
    then we can implement an $\epsilon$-approximated block encoding of $A^{-c}/(2\kappa_M^c)$ in complexity $\mathcal{O}( \kappa_M T_M (1+c) \log^2(  \frac{\kappa_M^{1+c}}{\epsilon} ) )$ where $T_M$ is the complexity to obtain the block encoding of $\mathcal{M}$. 
\end{lemma}
Since the matrix $A^T A$ is positive, we can apply the above lemma (with $\mathcal{M} = A^T A$ and $c= \frac{1}{2}$) to obtain the block encoding of $ \frac{1}{2\kappa_M^{c}} (A^T A)^{-c }  $. We mention the following spectral property. Let $\{ \lambda_i, \ket{\lambda_i} \}_{i=1}^n$ denotes the spectrum, including eigenvalues and corresponding eigenvectors of $A$, then $\{ \lambda_i^2, \ket{\lambda_i} \}_{i=1}^n$ is the spectrum of $A^T A$. Therefore, if $A$ is positive semidefinite, or $\lambda_i \geq 0$ for all $i=1,2,...,n$, then $\sqrt{\lambda_i^2} = \lambda_i$, so $ (A^T A)^{-1/2} = A^{-1}$. Additionally, if $\kappa$ is the conditional number of $A$, which is the ratio between the largest and smallest eigenvalue of $A$, then the conditional number $\kappa_M$ of $A^T A$ is $\kappa_M = \kappa^2$. Provided that we can prepare the state $\textbf{b} \equiv \ket{\textbf{b}}$ (assuming to have unit norm for convenience), e.g., via Lemma \ref{lemma: stateprepration}, we can then take the block encoding of $\frac{1}{2\kappa_M^{c}} (A^T A)^{-c }  = \frac{1}{2\kappa} A^{-1} $ and apply it to $\ket{\textbf{b}}$. According to definition \ref{def: blockencode}, we obtain the state:
\begin{align}
    \ket{\bf 0} \frac{1}{2\kappa}A^{-1} \ket{\textbf{b}} + \ket{\rm Garbage}
\end{align}
Measuring the ancilla and post-select on $\ket{\bf 0}$, we obtain the state $\varpropto A^{-1}\ket{\textbf{b}}$. The success probability of this measurement is $ \frac{1}{4\kappa^2}|| A^{-1} \ket{\textbf{b}}||^2 = \mathcal{O} \big( \frac{1}{4\kappa^2}\big) $, which can be improved quadratically faster using the amplitude amplification technique \cite{brassard2002quantum}. The complexity of approach is simply the product of the complexity of producing the block-encoded $A^T A$, of using Lemma \ref{lemma: negative} (with $c = 1/2$ and $\mathcal{M} = A^T A$), and of measuring at the final step to obtain $\ket{\xbf}$. Thus, the total complexity is is $\mathcal{O}\Big( \kappa^3 \log(s) \log^2 \frac{\kappa^{3/2}}{\epsilon}  \Big)$.

The above procedure works only when $A$ is positive semidefinite, because in such a case $ (A^T A)^{-1/2} = A^{-1} $. For a general $A$, it might not hold and we can modify the above algorithm as follows. As the eigenvalues $\{\lambda_i\}_{i=1}^n$ of $A$ are between $(-1,1)$, the shifted matrix $\frac{1}{2}\big( \Ibb_n + A\big)$ has eigenvalues $\{ \frac{1}{2}(1+\lambda_i)\}_{i=1}^n $ falling between $(0,1)$, which indicates that this matrix is positive semidefinite. It is also clear that the conditional number of this shifted matrix is upper bounded by $2$, which is very small. The matrix representation of this matrix is:
\begin{align}
    \frac{1}{2}\big( \Ibb_n + A\big) =\frac{1}{2} \begin{pmatrix}
        A_{11}+1 & A_{12} &  \cdots & A_{1n} \\
        A_{21} & A_{22}+1 & \cdots & A_{2n} \\
        \vdots & \vdots & \ddots & \vdots \\
        A_{n1} &  A_{n2} & \cdots & A_{nn}+1
    \end{pmatrix}
\end{align}
which is a slight adjustment of the original matrix $A$. Thus, by using the same procedure that we used to prepare the block encoding of $\mathcal{X}^T \mathcal{X}$ from the beginning of Section \ref{sec: quantumalgorithm}, we can obtain the block encoding of $\frac{1}{2}\big( \Ibb_n + A\big)^T \frac{1}{2}\big( \Ibb_n + A\big) $. Because $\frac{1}{2}\big( \Ibb_n + A\big) $ is positive semidefinite,  we have pointed out in the previous paragraph the property that $ \Big( \frac{1}{2}\big( \Ibb_n + A\big)^T \frac{1}{2}\big( \Ibb_n + A\big) \Big)^{1/2} = \frac{1}{2}\big( \Ibb_n + A\big) $. Thus, an application of the following lemma
\begin{lemma}[Positive Power Exponent \cite{gilyen2019quantum},\cite{chakraborty2018power}]
\label{lemma: positive}
    Given a block encoding of a positive matrix $\mathcal{M}$ such that 
    $$ \frac{\Ibb}{\kappa_M} \leq \mathcal{M} \leq \Ibb. $$
   Let $c \in (0,1)$. Then we can implement an $\epsilon$-approximated block encoding of $\mathcal{M}^c/2$ in time complexity $\mathcal{O}( \kappa_M T_M \log^2 (\frac{\kappa_M}{\epsilon})  )$, where $T_M$ is the complexity to obtain the block encoding of $\mathcal{M}$. 
\end{lemma}
with $\mathcal{M} =\frac{1}{2}\big( \Ibb_n + A\big)^T \frac{1}{2}\big( \Ibb_n + A\big)  $ and $c= \frac{1}{2}$ allows us to construct the block encoding of $ \frac{1}{4}\big( \Ibb_n + A\big)  $. As noted in the definition \ref{def: blockencode}, an identity matrix $\Ibb_n$ can be simply block-encoded, we can then use Lemma \ref{lemma: scale} to construct the block encoding of $ \frac{1}{4}\Ibb_n$. Then we use Lemma \ref{lemma: sumencoding} to construct the block encoding of:
\begin{align}
    \frac{1}{2} \Big( \frac{1}{4}\big( \Ibb_n + A\big) -  \frac{1}{4}\Ibb_n\Big) = \frac{A}{8}
\end{align}
Then applying Lemma \ref{lemma: negative} (with $c=1$) yields the block encoding of $\frac{1}{2\kappa} A^{-1}$, which can then be used to obtain the state $\varpropto A^{-1} \ket{\textbf{b}}$ as we discussed in the previous paragraph. 

To analyze the complexity, we recall that the complexity to produce the block encoding of $\frac{1}{2}\big( \Ibb_n + A\big)^T \frac{1}{2}\big( \Ibb_n + A\big)  $ is $\mathcal{O}\big( \log sn \big)$. Then we use Lemma \ref{lemma: positive} to transform it into $ \frac{1}{4}\big( \Ibb_n + A\big) $, and the complexity of this step is $\mathcal{O}\big(  \log(sn) \log^2 \frac{1}{\epsilon} \big)$ (we ignore the conditional number of $\frac{1}{2}\big( \Ibb_n + A\big)^T \frac{1}{2}\big( \Ibb_n + A\big) $ because we pointed out before that it is upper bounded by $2$, which is very small). Then we use Lemma \ref{lemma: sumencoding} to construct the block encoding of $ \frac{A}{8}$, which incurs a further $\mathcal{O}(1)$ cost because the block encoding of $\Ibb_n$ has $\mathcal{O}(1)$ cost (see def.~\ref{def: blockencode}), and given that Lemma \ref{lemma: sumencoding} use the block encoding of $  \frac{1}{4}\big( \Ibb_n + A\big)$ one time, so the complexity to produce $\frac{A}{8}$ is $ \mathcal{O}\big(  \log(sn) \log^2 \frac{1}{\epsilon} \big)$. The next step is using Lemma \ref{lemma: negative} (with $c= 1$ and $\mathcal{M} = A/8$) to transform the block-encoded $\frac{A}{8} $ into $ \frac{A^{-1}}{\kappa}$, which results in complexity:
$$  \mathcal{O}\Big( \kappa^2 \log(sn) \log^2\big( \frac{\kappa^2}{\epsilon}\big) \log^2 \frac{1}{\epsilon}  \Big)$$

We summarize the the result of this section in the following:
\begin{theorem}[Refined Quantum Linear Solver]
Let the linear system be $A\xbf = \textbf{b}$ where $A$ is an $s$-sparse, Hermitian matrix of size $n \times n$, with conditional number $\kappa$, and  $\textbf{b}$ is unit. Then there is a quantum algorithm outputting the state $\ket{\xbf } \varpropto  A^{-1}\textbf{b} $ in complexity
$$ \mathcal{O}\Big( \kappa^2 \log(sn) \log^2\big( \frac{\kappa^2}{\epsilon}\big) \log^2 \frac{1}{\epsilon}  \Big)$$
In the case $A$ is positive-semidefinite, the complexity is:
$$ \mathcal{O}\Big( \kappa^3 \log (sn) \log^2 \frac{\kappa^{3/2}}{\epsilon}  \Big)$$
\end{theorem}

\section{Application and Implication}
\label{sec: applicationandimplication}
We discuss a few applications and implications of the results as well as related techniques established in previous sections. \\

\noindent
\textbf{Direct quantum simulation.} As mentioned earlier, the key objective of quantum simulation is to (approximately) construct the evolution operator $\exp(-i Ht)$. In the above, we have shown how to obtain the block encoding of $\varpropto A$, provided that its columns are classically known. In a similar manner, if the columns of the Hamiltonian $H$ of interest are known, then we can follow the same procedure as above and construct the block encoding of $\varpropto H$, with complexity $\mathcal{O}\big( \log (n) \log^2 \frac{1}{\epsilon} \big)$. We note that similar to existing works, we assume the norm of $H$ is less than $1$. Otherwise, we can consider a rescaled Hamiltonian $\frac{H}{|H|_{\max}}$ where $|H|_{\max}$ is the maximum element of $H$, and then aim to simulate for a longer time $|H|_{\max} t$. To obtain the operator $\exp(-i H t)$, we can apply the results of \cite{low2017optimal,low2019hamiltonian, gilyen2019quantum}. More concretely, we leverage the Lemma \ref{lemma: theorem56} (in the appendix) and choose the polynomial $P$ to be an approximation of $\exp(-i H t)$ (Jacobi-Anger expansion). According to Theorem 58 of \cite{gilyen2019quantum}, this polynomial $P$ has degree $\mathcal{O}\Big( |H|_{\max} t + \frac{\log (1/\epsilon)}{\log\big(e + \log(1/\epsilon)/t)  \big)}  \Big)$. Per Lemma \ref{lemma: theorem56}, we can obtain the (block-encoded) simulation operator $\exp(-iH t)$ with complexity $\mathcal{O}\Big( \log (n) \log^2 \frac{1}{\epsilon}\big(   |H|_{\max} t + \frac{\log (1/\epsilon)}{\log\big(e + \log(1/\epsilon)/t)  \big)}  \big) \Big) $. As established, this is optimal with respect to time $t$ and dimension $n$, while being nearly optimal in the inverse of error tolerance. \\

\noindent
\textbf{Quantum simulation by solving linear equation.} The above method features a direct simulation, where we construct the evolution operator $\exp(-i H t)$ directly, leveraging the technique outlined in previous context. Here we consider an alternative, indirect way, which is reducing Schrodinger's equation into a linear equation, for which we can apply the result from previous section. This reduction strategy has been employed in many previous works \cite{childs2021high, berry2017quantum, childs2020quantum}, where the authors considered more general problems, including solving linear, nonlinear ordinary differential equations and partial differential equations. Recall that the Schrodinger's equation is a first-order ordinary differential equation $\frac{\partial \ket{\psi}}{\partial t} = -i H \ket{\psi}$. Dividing the time interval $[0,t]$ into subintervals $[0,\Delta, 2\Delta, ..., N\Delta \equiv t]$ and defining $\ket{\psi}_k = \big( \psi_1(k \Delta), \psi_2(k \Delta), ..., \psi_n(k \Delta) \big)^T$. A simple approximation of the derivative of, say, $\psi_j(k\Delta)$, reads $\frac{\partial \psi_j }{\partial t } |_{k\Delta} = \frac{1}{2\Delta}\big(  \psi_j(k\Delta+\Delta) - \psi_j(k \Delta-\Delta) \big) \longrightarrow \ket{\psi}_{k+1} - \ket{\psi}_{k-1} = (-2i\Delta) H \ket{\psi}_k$. For $k=0$, which is the starting point, we can use $\frac{\partial \psi_j }{\partial t } |_{0} = \frac{1}{\Delta}  \big( \psi_j (\Delta) - \psi_j(0)  \big)   $. So we obtain the following equations:
\begin{align}
    \begin{cases}
      \ket{\psi}_1 -   \ket{\psi}_0  = (-i\Delta) H \ket{\psi}_0 \\
        \ket{\psi}_{2} - \ket{\psi}_{0} = (-2i\Delta) H \ket{\psi}_1\\
        \ket{\psi}_{3} - \ket{\psi}_{1} = (-2i\Delta) H \ket{\psi}_2 \\
        \vdots \\
        \ket{\psi}_{N} - \ket{\psi}_{N-2} = (-2i\Delta) H \ket{\psi}_{N-1}
    \end{cases}
\end{align}
which forms a linear equation: 
\begin{align}
    \begin{pmatrix}
        i\Delta H& \Ibb_n & 0 & 0 & \cdots & 0 \\
        -\Ibb_n & 2i\Delta H & \Ibb_n & 0 & \cdots & 0 \\
        0 & -\Ibb_n & 2i\Delta H & \Ibb_n & \cdots & 0 \\
        0 & 0 & -\Ibb_n & 2i\Delta H & \cdots & 0 \\
        \vdots & \vdots & \vdots & \vdots & \ddots & \vdots \\
        0 & 0 & 0 & -\Ibb_n & 2i\Delta H & \Ibb_n 
    \end{pmatrix} \begin{pmatrix}
        \ket{\psi}_0 \\
        \ket{\psi}_1 \\
        \ket{\psi}_2  \\
        \ket{\psi}_3 \\
        \vdots  \\
        \ket{\psi}_N
    \end{pmatrix} \\ = \begin{pmatrix}
        \ket{\psi}_0 \\
        0 \\
        0  \\
        0 \\
        \vdots  \\
        0
    \end{pmatrix}
\end{align}
This is a linear system of size $ nN \times nN$, and apparently we can use our refined quantum linear solver to find the state $\varpropto \sum_{k=0}^N \ket{k} \ket{\psi}_k$. The above method used a simple approximation for the derivative, thus reducing the Schrodinger's equation, which is an ODE to a linear equation. We remind that this strategy was already used in \cite{berry2014high} to solve ordinary differential equations. According to them, a more advanced method, namely, general linear multistep method, yields the following equation at each time step $k\Delta$: $\sum_{l=-K}^K \alpha_l \ket{\psi}_{k+l} =  (-i\Delta) \sum_{l=-K}^K H \beta_l \ket{\psi}_{k+l} $. We remark that for $k < K$, $k-K < 0$ therefore we can't use the multistep, instead, we use the approximation as above $\ket{\psi}_{k+1} - \ket{\psi}_{k-1} = (-2i\Delta) H \ket{\psi}_k $ and only use multisteps for $k\geq K$. We thus form an equation:
\begin{align}
\begin{cases}
     \ket{\psi}_1 -   \ket{\psi}_0  = (-i\Delta) H\ket{\psi}_0 \\
        \ket{\psi}_{2} - \ket{\psi}_{0} = (-2i\Delta) H \ket{\psi}_1\\
        \ket{\psi}_{3} - \ket{\psi}_{1} = (-2i\Delta) H \ket{\psi}_2 \\
        \vdots \\
        \ket{\psi}_{K} - \ket{\psi}_{K-2} = (-2i\Delta) H \ket{\psi}_{K-1} \\
    \sum_{l=-K}^K \alpha_l \ket{\psi}_{K+l} =  (-i\Delta) \sum_{l=-K}^K H \beta_l \ket{\psi}_{K+l} \\
    \sum_{l=-K}^K \alpha_l \ket{\psi}_{K+1+l} =  (-i\Delta) \sum_{l=-K}^K H \beta_l \ket{\psi}_{K+1+l} \\
    \sum_{l=-K}^K \alpha_l \ket{\psi}_{K+2+l} =  (-i\Delta) \sum_{l=-K}^K H \beta_l \ket{\psi}_{K+2+l} \\
    \vdots \\
    \sum_{l=-K}^K \alpha_l \ket{\psi}_{N-K+l} =  (-i\Delta) \sum_{l=-K}^K H \beta_l \ket{\psi}_{N-K+l} \Big) 
\end{cases}
\label{29}
\end{align}
and thus form a more complicated linear systems. We refer to \cite{berry2014high} for a more detailed representation of this linear system. Because we discretize the time interval and approximate the derivation, there is an error induced, which means that each of state $\ket{\psi}_1,\ket{\psi}_2,..., \ket{\psi}_N$ above has some deviation to the true solution of the original differential equations, at corresponding time step. According to \cite{berry2014high} (see their Section IV), by choosing $N = \mathcal{O}\Big(  \frac{t^{1+1/K}}{\epsilon^{1/K}} \Big) $, then the accumulated error is $\epsilon$, i.e., for all $k=1,2,...,N$, we have that $\big|\big| \ket{\psi}_k - \ket{\psi}_{k}^{\rm true}    \big|\big| \leq \epsilon$, where $\ket{\psi}_{k}^{\rm true}  $ denotes the true solution. In addition, Theorem 7 of \cite{berry2014high} shows that the conditional number of the above system is $\mathcal{O}(N)$. Therefore, an application of our quantum linear solver yields a quantum simulation algorithm with complexity $\mathcal{O}\Big( \kappa^2 \log (nN) \log^2 \big( \frac{\kappa^{2}}{\epsilon}\big) \log^2 \frac{1}{\epsilon}  \Big) = \mathcal{\Tilde{O}}\Big(  \frac{t^{2+2/K}}{\epsilon^{2/K}} \log(n)   \Big)$, where $\mathcal{\Tilde{O}}$ hides the polylogarithmic terms. 

This approach is clearly not as efficient as the direct simulation approach above, especially with respect to time $t$ and inverse of error $1/\epsilon$. However, it does have some implications. First, we recall that in the original quantum linear solving algorithm \cite{harrow2009quantum} (HHL algorithm), the authors proved that the complexity on conditional number $\kappa$ cannot be better than linear, i.e., a sublinear scaling $\kappa^{1-\gamma}$ is not possible. Here, we provide an alternative and much simpler proof to this statement, based on the fact that our quantum simulation method uses a quantum linear solver as a subroutine. We recall from the above that the conditional number of the linear system defined in Eqn.~\ref{29} is $\kappa = \mathcal{O}(N)$, and the value of $N$ (the number of time steps) is $N \varpropto t^{1+1/K}$, which is sublinear. Therefore, $\kappa \varpropto t^{1+1/K} $. If a quantum linear solver can produce the solution in $\kappa^{1-\gamma}$, it means that it can solve Eqn.~\ref{29}, which encodes the evolved state at the time $t = N\Delta$, in complexity $\kappa^{1-\gamma} = N^{1-\gamma} = \big( t^{1+1/K} \big)^{1-\gamma} $. By choosing $\gamma$ properly, then $\big( t^{1+1/K} \big)^{1-\gamma} $ can be sublinear in $t$, which means that we can simulate the dynamics of a given quantum system in sublinear time. This violates the well-known no-forwarding theorem, which states that the complexity of simulating quantum system is $\Omega(t)$. Therefore, a quantum algorithm for solving linear system cannot have sublinear scaling in $\kappa$. 

Second, this approach can be extended to time-dependent regime in a straightforward manner, meanwhile the direct approach above cannot. In the time-dependent regime, the Hamiltonian $H$ becomes time-dependent, and we need to modify the linear system by setting $H$ (in Eqn.~\ref{29}) with $H_{k\Delta}$ -- which is the Hamiltonian at $k$-th time step. More specifically, we obtain the following:
\begin{align}
\begin{cases}
     \ket{\psi}_1 -   \ket{\psi}_0  = (-i\Delta) H_0\ket{\psi}_0 \\
        \ket{\psi}_{2} - \ket{\psi}_{0} = (-2i\Delta) H_1 \ket{\psi}_1\\
        \ket{\psi}_{3} - \ket{\psi}_{1} = (-2i\Delta) H_2 \ket{\psi}_2 \\
        \vdots \\
        \ket{\psi}_{K} - \ket{\psi}_{K-2} = (-2i\Delta) H_{K-1} \ket{\psi}_{K-1} \\
    \sum_{l=-K}^K \alpha_l \ket{\psi}_{K+l} =  (-i\Delta) \sum_{l=-K}^K H_{K+l} \beta_l \ket{\psi}_{K+l} \\
    \sum_{l=-K}^K \alpha_l \ket{\psi}_{K+1+l} =  (-i\Delta) \sum_{l=-K}^K H_{K+1+l} \beta_l \ket{\psi}_{K+1+l} \\
    \sum_{l=-K}^K \alpha_l \ket{\psi}_{K+2+l} =  (-i\Delta) \sum_{l=-K}^K H_{K+2+l} \beta_l \ket{\psi}_{K+2+l} \\
    \vdots \\
    \sum_{l=-K}^K \alpha_l \ket{\psi}_{N-K+l}  =  (-i\Delta) \sum_{l=-K}^K H_{N-K+l} \beta_l \ket{\psi}_{N-K+l} 
\end{cases}
\end{align}
Solving this linear equation yields the state $\varpropto \sum_{k=0}^N \ket{k} \ket{\psi}_k$ -- which includes the evolved states at different time step, from $0$ up to $N \Delta = t$.

\section{Outlook and Conclusion}
In this work, we have described two refinements of the quantum algorithm for principal component analysis and for solving linear algebraic equations. Our algorithms are largely motivated by the caveats faced by existing methods, which were identified and improved in our work. More specifically, for the PCA, we have pointed out that prior constructions suffered from both strong input assumption and poor scaling in certain parameters, which severely limit their impact and potential realization, to some extent. We introduced two alternative approaches for performing PCA. The first one is making use of the power method, which is a simple yet highly efficient tool for dealing with top eigenvalues/eigenvectors. As we have seen, the top eigenvalues/eigenvectors, also called the principal components of the covariance matrix, could be revealed within (poly)logarithmic complexity in all parameters. This approach surpasses the previous results \cite{lloyd2014quantum} and \cite{nghiem2025new} in terms of complexity scaling in the inverse of error tolerance. However, as we discussed, it is most effective when the gap between the largest eigenvalues of the covariance matrix is sufficiently large. The second approach, on the other hand, relies on gradient descent algorithm, and its performance does not depend on the gap, with the trade-off of having linear scaling in inverse of error tolerance. Therefore, these two approaches can complement each other in practice. We then extend the technique from PCA to the context of solving linear equations and show that a highly efficient quantum linear solver can be achieved. The complexity turns out to scale (poly)logarithmically in most parameters, except the conditional number. This is an exponential improvement over the previous results \cite{harrow2009quantum, childs2017quantum, nghiem2025new2, clader2013preconditioned}. In particular, we have shown that the techniques of our new QPCA/QLSA can be used in a direct manner in the context of quantum simulation. The result is a new input model where efficient quantum simulation is possible \cite{zhang2022quantum, gilyen2019quantum}. In addition, based on the reduction from the simulation problem to the problem of solving linear equations, we showed that the quantum algorithm cannot invert a matrix in sublinear time $\kappa^{1-\gamma}$, which provides a simpler proof for the same result that already appeared in \cite{harrow2009quantum}. The probably most important aspect of our new QCPA, QLSA and quantum simulation algorithms is that they do not require oracle/black-box access to classical data, which eases a significant amount of hardware constraints, suggesting a great potential for experimental realization.

\section*{Acknowledgement}
We acknowledge support from Center for Distributed Quantum Processing, Stony Brook University.

\bibliography{ref.bib}{}
\bibliographystyle{unsrt}

\onecolumngrid
\appendix
\section{Preliminaries}
\label{sec: prelim}
Here, we summarize the main recipes of our work, mostly derived from the seminal QSVT work \cite{gilyen2019quantum}. We keep the statements brief and precise for simplicity, with their proofs/ constructions referred to in their original works.

\begin{definition}[Block Encoding Unitary]~\cite{low2017optimal, low2019hamiltonian, gilyen2019quantum}
\label{def: blockencode} 
Let $A$ be some Hermitian matrix of size $N \times N$ whose matrix norm $|A| < 1$. Let a unitary $U$ have the following form:
\begin{align*}
    U = \begin{pmatrix}
       A & \cdot \\
       \cdot & \cdot \\
    \end{pmatrix}.
\end{align*}
Then $U$ is said to be an exact block encoding of matrix $A$. Equivalently, we can write $U = \ket{ \bf{0}}\bra{ \bf{0}} \otimes A + (\cdots)$, where $\ket{\bf 0}$ refers to the ancilla system required for the block encoding purpose. In the case where the $U$ has the form $ U  =  \ket{ \bf{0}}\bra{ \bf{0}} \otimes \Tilde{A} + (\cdots) $, where $|| \Tilde{A} - A || \leq \epsilon$ (with $||.||$ being the matrix norm), then $U$ is said to be an $\epsilon$-approximated block encoding of $A$. Furthermore, the action of $U$ on some quantum state $\ket{\bf 0}\ket{\phi}$ is:
\begin{align}
    \label{eqn: action}
    U \ket{\bf 0}\ket{\phi} = \ket{\bf 0} A\ket{\phi} +  \ket{\rm Garbage},
\end{align}
where $\ket{\rm Garbage }$ is a redundant state that is orthogonal to $\ket{\bf 0} A\ket{\phi}$. The above definition has multiple natural \textbf{corollaries}: 
\begin{itemize}
    \item First, an arbitrary unitary $U$ block encodes itself
    \item Second, suppose that $A$ is block encoded by some matrix $U$, then $A$ can be block encoded in a larger matrix by simply adding any ancilla (supposed to have dimension $m$), then note that $\Ibb_m \otimes U$ contains $A$ in the top-left corner, which is block encoding of $A$ again by definition 
    \item Third, it is almost trivial to block encode identity matrix of any dimension. For instance, we consider $\sigma_z \otimes \Ibb_m$ (for any $m$), which contains $\Ibb_m$ in the top-left corner. 
\end{itemize}
\end{definition}



\begin{lemma}[Block Encoding of Product of Two Matrices]
\label{lemma: product}
    Given the unitary block encoding of two matrices $A_1$ and $A_2$, then there exists an efficient procedure that constructs a unitary block encoding of $A_1 A_2$ using each block encoding of $A_1,A_2$ one time. 
\end{lemma}

\begin{lemma}[\cite{camps2020approximate} \revise{Block Encoding of a Tensor Product}]
\label{lemma: tensorproduct}
    Given the unitary block encoding $\{U_i\}_{i=1}^m$ of multiple operators $\{M_i\}_{i=1}^m$ (assumed to be exact encoding), then, there is a procedure that produces the unitary block encoding operator of $\bigotimes_{i=1}^m M_i$, which requires \revise{parallel single uses} of 
    $\{U_i\}_{i=1}^m$ and $\mathcal{O}(1)$ SWAP gates. 
\end{lemma}
The above lemma is a result in \cite{camps2020approximate}. 
\begin{lemma}[\revise{\cite{gilyen2019quantum} Block Encoding of a  Matrix}]
\label{lemma: As}
    Given oracle access to $s$-sparse matrix $A$ of dimension $n\times n$, then an $\epsilon$-approximated unitary block encoding of $A/s$ can be prepared with gate/time complexity $\mathcal{O}\Big(\log n + \log^{2.5}(\frac{s^2}{\epsilon})\Big).$
\end{lemma}
This is presented in~\cite{gilyen2019quantum} (see their Lemma 48), and one can also find a review of the construction in~\cite{childs2017lecture}. We remark further that the scaling factor $s$ in the above lemma can be reduced by the preamplification method with further complexity $\mathcal{O}({s})$~\cite{gilyen2019quantum}.

\begin{lemma}[\cite{gilyen2019quantum} Linear combination of block-encoded matrices]
    Given unitary block encoding of multiple operators $\{M_i\}_{i=1}^m$. Then, there is a procedure that produces a unitary block encoding operator of \,$\sum_{i=1}^m \pm M_i/m $ in complexity $\mathcal{O}(m)$, e.g., using block encoding of each operator $M_i$ a single time. 
    \label{lemma: sumencoding}
\end{lemma}

\begin{lemma}[Scaling Block encoding] 
\label{lemma: scale}
    Given a block encoding of some matrix $A$ (as in~\ref{def: blockencode}), then the block encoding of $A/p$ where $p > 1$ can be prepared with an extra $\mathcal{O}(1)$ cost.  
\end{lemma}
To show this, we note that the matrix representation of RY rotational gate is
\begin{align}
   R_Y(\theta) = \begin{pmatrix}
        \cos(\theta/2) & -\sin(\theta/2) \\
        \sin(\theta/2) & \cos(\theta/2) 
    \end{pmatrix}.
\end{align}
If we choose $\theta$ such that $\cos(\theta/2) = 1/p$, then Lemma~\ref{lemma: tensorproduct} allows us to construct block encoding of $R_Y(\theta) \otimes \mathbb{I}_{{\rm dim}(A)}$  (${\rm dim}(A)$ refers to dimension of matirx $A$), which contains the diagonal matrix of size ${\rm dim}(A) \times {\rm dim}(A)$ with entries $1/p$. Then Lemma~\ref{lemma: product} can construct block encoding of $(1/p) \ \mathbb{I}_{{\rm dim}(A)} \cdot A = A/p$.  \\

The following is called amplification technique:
\begin{lemma}[\cite{gilyen2019quantum} Theorem 30; \revise{\bf Amplification}]\label{lemma: amp_amp}
Let $U$, $\Pi$, $\widetilde{\Pi} \in {\rm End}(\mathcal{H}_U)$ be linear operators on $\mathcal{H}_U$ such that $U$ is a unitary, and $\Pi$, $\widetilde{\Pi}$ are orthogonal projectors. 
Let $\gamma>1$ and $\delta,\epsilon \in (0,\frac{1}{2})$. 
Suppose that $\widetilde{\Pi}U\Pi=W \Sigma V^\dagger=\sum_{i}\varsigma_i\ket{w_i}\bra{v_i}$ is a singular value decomposition. 
Then there is an $m= \mathcal{O} \Big(\frac{\gamma}{\delta}
\log \left(\frac{\gamma}{\epsilon} \right)\Big)$ and an efficiently computable $\Phi\in\mathbb{R}^m$ such that
\begin{equation}
\left(\bra{+}\otimes\widetilde{\Pi}_{\leq\frac{1-\delta}{\gamma}}\right)U_\Phi \left(\ket{+}\otimes\Pi_{\leq\frac{1-\delta}{\gamma}}\right)=\sum_{i\colon\varsigma_i\leq \frac{1-\delta}{\gamma} }\tilde{\varsigma}_i\ket{w_i}\bra{v_i} , \text{ where } \Big|\!\Big|\frac{\tilde{\varsigma}_i}{\gamma\varsigma_i}-1 \Big|\!\Big|\leq \epsilon.
\end{equation}
Moreover, $U_\Phi$ can be implemented using a single ancilla qubit with $m$ uses of $U$ and $U^\dagger$, $m$ uses of C$_\Pi$NOT and $m$ uses of C$_{\widetilde{\Pi}}$NOT gates and $m$ single qubit gates.
Here,
\begin{itemize}
\item C$_\Pi$NOT$:=X \otimes \Pi + I \otimes (I - \Pi)$ and a similar definition for C$_{\widetilde{\Pi}}$NOT; see Definition 2 in \cite{gilyen2019quantum},
\item $U_\Phi$: alternating phase modulation sequence; see Definition 15 in \cite{gilyen2019quantum},
\item $\Pi_{\leq \delta}$, $\widetilde{\Pi}_{\leq \delta}$: singular value threshold projectors; see Definition 24 in \cite{gilyen2019quantum}.
\end{itemize}
\end{lemma}

\begin{lemma}[Projector]
\label{lemma: projector}
The block encoding of a projector $\ket{j-1}\bra{j-1}$ (for any $j=1,2, ...,n$) by a circuit of depth $\mathcal{O}\big( \log n \big)$ 
\end{lemma}
\noindent
\textit{Proof.} First we note that it takes a circuit of depth $\mathcal{O}(1)$ to generate $\ket{j-1}$ from $\ket{0}$. Then Lemma \ref{lemma: improveddme} can be used to construct the block encoding of $\ket{j-1}\bra{j-1}$. 
\begin{lemma}[\cite{guo2024nonlinear}, or Theorem 2 in \cite{rattew2023non}]
\label{lemma: diagonal}
     Given an n-qubit quantum state specified by a state-preparation-unitary $U$, such that $\ket{\psi}_n=U\ket{0}_n=\sum^{N-1}_{k=0}\psi_k \ket{k}_n$ (with $\psi_k \in \mathbb{C}$ and $N=2^n$), we can prepare an exact block-encoding $U_A$ of the diagonal matrix $A = {\rm diag}(\psi_0, ...,\psi_{N-1})$ with $\mathcal{O}(n)$ circuit depth and a total of $\mathcal{O}(1)$ queries to a controlled-$U$ gate  with $n+3$ ancillary qubits.
\end{lemma}

\section{More Details on prior QPCA algorithms}
\label{sec: moredetailspca}
Here we provide more technical details of the discussion in Section \ref{sec: overviewPCA}, where we mention previous progress regarding QPCA, specifically \cite{lloyd2014quantum, nghiem2025new}. \\

\noindent
\textbf{Ref.~\cite{lloyd2014quantum}.} This work's initial motivation was actually simulating density matrix, i.e., obtaining $\exp(- i \rho t)$ from multiple copies of density state $\rho \in \mathbb{C}^{n \times n}$ where $n$ is the dimension. In order to obtain the unitary transformation $\exp(- i \rho t)$, the authors in \cite{lloyd2013quantum} used the following property:
\begin{align}
    \Tr_1 \exp(-iS \Delta t) \big( \rho \otimes \sigma \big) \exp(-iS \Delta t) = \sigma - i \Delta t [\rho,\sigma] + \mathcal{O}(\Delta t^2) \approx \exp(-i \rho \Delta t) \sigma \exp(-i \rho \Delta t)
\end{align}
where $\Tr_1$ is the partial trace over the first system, $S$ is the swap operator between two system of $\log(n)$ qubits, and $\sigma$ is some ancilla system. Defining $\exp(-i \rho \Delta t) \sigma \exp(-i \rho \Delta t) = \rho_1 $. Repeat the above step:
\begin{align}
     \Tr_1 \exp(-iS \Delta t)  \big( \rho \otimes \rho_1   \big) \exp(-iS \Delta t) &\approx \exp(-i\rho \Delta t) \big( \exp(-i \rho \Delta t) \sigma \exp(-i \rho \Delta t) \big)  \exp(-i \rho \Delta t) \\
     &= \exp(- i\rho 2\Delta t) \sigma \exp(-i \rho 2\Delta t) 
\end{align}
To obtain $\exp(-i \rho t)$, we repeat the above procedure $N$ times, then we obtain:
\begin{align}
    \exp(-i \rho N \Delta t) \sigma \exp(-i \rho N \Delta t)
\end{align}
The authors in \cite{lloyd2013quantum} shows that to simulate $\exp(-i \rho t)$  to accuracy $\epsilon$, then it requires:
\begin{align}
    N = \mathcal{O} \big(  \frac{t^2}{\epsilon}  \big)
\end{align}
total number of copies and repetition, where $t = N\Delta t$. To find the top eigenvalues/eigenvectors, the authors in \cite{lloyd2014quantum} used quantum phase estimation with $\rho$ as input state. Denote the spectrum of $\rho$ as $\{ \alpha_i , \ket{\phi_i} \}$. The outcome of phase estimation algorithm is a density state:
\begin{align}
    \sum_{i} \alpha_i \ket{\Tilde{\alpha}_i} \bra{ \Tilde{\alpha}_i} \otimes \ket{\phi_i}\bra{\phi_i}
\end{align}
where $ \Tilde{\alpha}_i $ is a binary string approximation of $\alpha_i$. By sampling from the above state, we can obtain the highest eigenvalues / eigenvectors because the probability to obtain the highest eigenvalues is $|\alpha_i|^2$, which means that the higher the value, the higher probability. According to \cite{lloyd2014quantum}, in order to guarantee that the error of eigenvalues estimation is $\epsilon$, we need to choose $t = \mathcal{O}(1/\epsilon^2) $. So the total complexity of this approach is $ \mathcal{O}(1/\epsilon^3)$. 

To apply this approach in the context of principal component analysis, the Ref.~\cite{lloyd2014quantum} assumed that, via some oracle (or quantum random access memory), the ability to prepare a density state $\rho \varpropto \mathcal{C}$ (where $\mathcal{C}$ is the covariance matrix) in logarithmic time $\mathcal{O}(\log mn)$. Then the above procedure yields the top $r$ eigenvalues/ eigenvectors with complexity $\mathcal{O}\Big(  r \frac{1}{\epsilon^3} \log mn  \Big)  $, as one needs to repeat the sampling roughly $r$ times to obtain $r$  different eigenvalues/ eigenvectors. \\

\noindent
\textbf{Ref.~\cite{nghiem2025new}.} The approach of this work is a combination of the density matrix exponentiation technique above and the power method, which was also used in our main text. Instead of using $\exp(-i\rho t)$ with the phase estimation algorithm, the authors of \cite{nghiem2025new} leveraged the following result from \cite{gilyen2019quantum}:
\begin{lemma}[Logarithmic of Unitary, Corollay 71 in \cite{gilyen2019quantum}]
\label{lemma: logarithmicofunitary}
    Suppose that $U = \exp(-iH)$, where $H$ is a Hamiltonian of norm at most $1/2$. Let $\epsilon \in (0,1/2]$, then we can implement an $\epsilon$-approximated block encoding of $\pi H /2$ (see further definition \ref{def: blockencode}) with $\mathcal{O}(\log(\frac{1}{\epsilon} ))$ uses of controlled-U and its inverse, using $\mathcal{O}(\log(\frac{1}{\epsilon}))$ two-qubit gates and using a single ancilla qubit. 
\end{lemma}
The above lemma allows us to construct the block encoding of $\frac{\pi}{4}\rho$ from $\exp(-i \rho t)$ (by setting $t= 1/2)$. To prepare a covariance matrix without resorting on oracle/QRAM, we recall from the main text that the dataset contains $m$ samples $\xbf^1,\xbf^2, ..., \xbf^m$ where each $\xbf^i \in \Rbb^n$. In the context of \cite{gordon2022covariance} and \cite{nghiem2025new}, they assumed that each data is normalized, i.e., $ ||\xbf^i ||  =1$. Provided $\xbf^i$ is known, the amplitude encoding method \cite{grover2000synthesis,grover2002creating,plesch2011quantum, schuld2018supervised, nakaji2022approximate,marin2023quantum,zoufal2019quantum,prakash2014quantum, zhang2022quantum} can be used to prepare it with an efficient circuit $U_i$ of depth $\mathcal{O}(\log n)$. Suppose that from $\{ \xbf^1,\xbf^2, ..., \xbf^m \} $, we randomly select $\xbf^i$ with probability $1/m$, then we obtain an ensemble $\frac{1}{m} \sum_{i=1}^n \xbf^i (\xbf^i)^\dagger$. Using the above procedure, first simulate $\exp\big(-i \frac{1}{2m} \sum_{i=1}^n \xbf^i (\xbf^i)^\dagger \big)$ (with complexity $\mathcal{O}\big(  \frac{1}{\epsilon}\log n \big)$, then apply Lemma \ref{lemma: logarithmicofunitary} to construct the block encoding of $\frac{\pi}{4} \frac{1}{m} \sum_{i=1}^n \xbf^i (\xbf^i)^\dagger$. The resultant complexity is then $\mathcal{O}\big( \frac{1}{\epsilon}\log(\frac{1}{\epsilon}) \log n \big) $.

To construct the block encoding of $\frac{1}{m}\mu \mu^\dagger$, they use Lemma \ref{lemma: sumencoding} to construct the block encoding of $ \frac{1}{m} \sum_{i=1}^m U_i$, which contains $\frac{1}{m} \sum_{i}\xbf_i $ as the first column. The complexity of this step is $\mathcal{O}(m \log n)$ because each $U_i$ is used one time. Then they use Lemma \ref{lemma: improveddme} to construct the block encoding of $\Big( \frac{1}{m} \sum_{i}\xbf_i\Big) \Big( \frac{1}{m} \sum_{i}\xbf_i\Big)^\dagger \equiv \mu \mu^\dagger $, which can be combined with Lemma \ref{lemma: scale} to transform it to $\frac{\pi}{4} \mu \mu^\dagger $. Recall that covariance matrix $\mathcal{C} $ can be expressed as:
\begin{align}
    \mathcal{C} = \frac{1}{m} \sum_{i=1}^n \xbf^i (\xbf^i)^\dagger - \mu\mu^\dagger
\end{align}
Thus one can use the block encoding of $\frac{\pi}{4} \frac{1}{m} \sum_{i=1}^n \xbf^i (\xbf^i)^\dagger,\frac{\pi}{4} \mu \mu^\dagger  $ and Lemma \ref{lemma: sumencoding} to construct the block encoding of $ \frac{1}{2} \Big(\frac{\pi}{4} \frac{1}{m} \sum_{i=1}^n \xbf^i (\xbf^i)^\dagger- \frac{\pi}{4} \mu \mu^\dagger   \Big) $, which is $\frac{\pi}{8} \mathcal{C}$. The complexity of this method is $\mathcal{O}\big( \frac{1}{\epsilon}\log(\frac{1}{\epsilon}) \log n + m \log n  \big)$. 

Another method for preparing the covariance matrix, as provided in \cite{nghiem2025new}, is to use $U_i$ with Lemma \ref{lemma: improveddme} to construct the block encoding of $\xbf^i (\xbf^i)^\dagger$ for all $i=1,2,...,m$. Then one uses Lemma \ref{lemma: sumencoding} to construct the block encoding of $ \frac{1}{m} \sum_{i=1}^m \xbf^i (\xbf^i)^\dagger $. This construction has complexity $\mathcal{O}(m \log n)$. Given that the block encoding of $\mu \mu^\dagger $ is provided above, one can use Lemma \ref{lemma: sumencoding} to construct the block encoding of $ \frac{1}{2} \Big(\frac{1}{m} \sum_{i=1}^m \xbf^i (\xbf^i)^\dagger- \mu \mu^\dagger \Big) \equiv \frac{1}{2}  \mathcal{C} $, with total complexity $\mathcal{O}(m \log n)$. Then one can find the top eigenvector/eigenvalue of $\frac{\pi}{4}\rho$ through  Lemma \ref{lemma: largestsmallest}. From such an eigenstate, one repeat the above procedure: using copies of $\ket{\lambda_i}\bra{\lambda_i}$ and simulate $\exp(- i\ket{\lambda_i}\bra{\lambda_i} /2)  $, then use Lemma \ref{lemma: logarithmicofunitary} to recover $ \frac{\pi}{4}\ket{\lambda_i}\bra{\lambda_i}$. Then one considers finding the maximum eigenvalue/eigenvector of $\mathcal{C } - \lambda_1\ket{\lambda_1}\bra{\lambda_1}$, and continue this process for $r$ eigenvalues/eigenvectors. According to the analysis provided in \cite{nghiem2025new}, the circuit complexity for producing top $r$ eigenvalues/eigenvectors is $\mathcal{O }\Big( m \log(n) \big( \frac{1}{\Delta^2} \log^3 (\frac{n}{\epsilon} ) \frac{1}{\epsilon^2} \big)^r  \Big)$ where $\Delta$ is the gap between two largest eigenvalues. \\

\section{More Details on Prior Quantum Linear Solving Algorithms}
\label{sec: moredetaillinearsolver}
With similar purpose to the previous section, in the following, we provide more details about existing quantum linear solving algorithms. \\

\noindent
\textbf{Ref.~\cite{harrow2009quantum}.} Under the same notations and conditions as in Sec.~\ref{sec: solvinglinearequation}, with a further assumption that there is an oracle/black-box access to entries of $A$ (in an analogous manner to previous simulation contexts \cite{berry2007efficient, aharonov2003adiabatic, berry2012black}), this work first leveraged these simulation algorithms to perform $\exp(-i A t)$. Then they perform the quantum phase estimation with $\exp(-i At)$ and $\ket{\textbf{b}}$ as input state, to obtain:
\begin{align}
    \sum_{i=1}^n \beta_i \ket{\phi_i} \ket{\lambda_i}
\end{align}
where $\{ \lambda_i, \ket{\phi_i} \}$ is eigenvalues/eigenvectors of $A$ and $\{ \beta_i\}$ is the expansion coefficients of $\ket{\textbf{b}}$ in this basis, i.e., $ \ket{\textbf{b}} = \sum_{i=1}^n \beta_i \ket{\phi_i}$. Then they append an ancilla initialized in $\ket{0}$, and rotate the ancilla conditioned on the phase register:
\begin{align}
     \sum_{i=1}^n \beta_i \ket{\phi_i} \ket{\lambda_i} \ket{0} \longrightarrow  \sum_{i=1}^n \beta_i \ket{\phi_i} \ket{\lambda_i}\Big( \frac{1}{\kappa \lambda_i} \ket{0} + \sqrt{1- \frac{1}{\kappa^2 \lambda_i^2}}\ket{1} \Big)
\end{align}
By uncomputing the phase register, or reverse the phase estimation algorithm, and discard that register, we obtain: 
\begin{align}
    \sum_{i=1}^n \beta_i \ket{\phi_i} \Big( \frac{1}{\kappa \lambda_i} \ket{0} + \sqrt{1- \frac{1}{\kappa^2 \lambda_i^2}}\ket{1} \Big)
\end{align}
Measuring the ancilla and post-select on $\ket{0}$, we obtain a state $\varpropto  \sum_{i=1}^n \frac{\beta_i}{\kappa \lambda_i} \ket{\phi_i} = \frac{1}{\kappa} A^{-1} \ket{\textbf{b}} $. The complexity of this algorithm, as analyzed in \cite{harrow2009quantum}, is $\mathcal{\Tilde{O}}\big( \kappa^2 s^2 \log (n) \frac{1}{\epsilon}\big)$ where $\Tilde{O}$ hides the polylogarithmic factor. \\

\noindent
\textbf{Ref.~\cite{childs2017quantum}.} The above HHL algorithm makes use of a quantum phase estimation algorithm, which leads to an unavoidable scaling in $1/\epsilon$. The work of \cite{childs2017quantum} improves upon this aspect by making use of the following approximations: 
\begin{align}
  \text{\rm Fourier approximation:} \   A^{-1} \approx \sum_{j=1}^K \alpha_j \exp(-i A \Delta_j) \\
   \text{\rm Chebyshev approximation:} \ A^{-1} \approx \sum_{j=1}^K \alpha_j T_{j} (A)
\end{align}
By using more precise simulation algorithms \cite{berry2015hamiltonian, berry2015simulating}, the terms $\exp(-i A\Delta_j)$ can be approximated more efficiently. Implementation of Chebyshev polynomials is also known to be efficient via quantum walk technique \cite{childs2010relationship, berry2012black}. The summation $\sum_{j=1}^K \alpha_j \exp(-i A \Delta_j) $, $\sum_{j=1}^K \alpha_j T_{j} (A) $ can be constructed using the technique called linear combination of unitaries \cite{berry2015simulating}. The value of $K$ turns out to be $\mathcal{O}\big( \kappa^2 \log^2 \frac{\kappa}{\epsilon}\big)$. Overall, as provided in Theorem 3 and 4 of \cite{childs2017quantum}, the complexity for constructing $A^{-1}$ and eventually, obtaining $\varpropto A^{-1} \ket{\textbf{b}}$ is $\mathcal{O}\Big( s \kappa^2 \log^{2.5} \big(  \frac{\kappa}{\epsilon}\big)  \big(  \log n + \log^{2.5}\frac{\kappa}{\epsilon} \big)   \Big) $ and $\mathcal{O}\Big( s \kappa^2 \log^{2} \big(  \frac{\kappa}{\epsilon}\big)  \big(  \log n + \log^{2.5}\frac{\kappa}{\epsilon} \big)   \Big) $ for Fourier approximation approach and Chebyshev approximation approach, respectively. \\

\noindent
\textbf{Ref.~\cite{nghiem2025new2}.} This recently introduced approach for solving linear equations is based on reducing the original problem to an optimization problem, which can be solved by gradient descent. More specifically, given a linear system $A\xbf = \textbf{b}$, one can find $\xbf$ by minimizing the following function:
\begin{align}
    f(\xbf) = \frac{1}{2}||\xbf||^2 + \frac{1}{2} || A\xbf - \textbf{b}||^2
\end{align}
This strategy was also used in \cite{huang2019near} to solve linear system, however, they developed a variational algorithm and thus their algorithm is heuristic. The above formulation allows us to use the gradient descent algorithm to find the minima. As the above function is strongly convex, a global minima is also local minima, and thus convergence to such a minima is guaranteed. The gradient descent algorithm works by first initializing a random  vector $\xbf_0$, then iterate the following procedure $T$ times:
\begin{align}
    \xbf \leftarrow \xbf - \eta \bigtriangledown f(\xbf)
\end{align}
where $\eta$ is the learning hyperparmeter. In \cite{nghiem2025new2}, the author performed an embed $\xbf \longrightarrow \xbf \xbf^\dagger$, and in this new framework, the gradient descent's update rule is redefined as:
\begin{align}
    \xbf \xbf^\dagger \leftarrow  \big( \xbf - \eta \bigtriangledown f(\xbf)\big) \big( \xbf - \eta \bigtriangledown f(\xbf)\big)^\dagger 
\end{align}
which turns out to be $ \xbf \xbf^\dagger  - \eta \xbf \bigtriangledown^\dagger f(\xbf) - \eta \bigtriangledown f(\xbf)  \xbf^\dagger + \eta^2 \bigtriangledown f(\xbf) \bigtriangledown^\dagger f(\xbf)$. The gradient of $f(\xbf)$ is:
\begin{align}
    \bigtriangledown f(\xbf) = \xbf  + A^\dagger A \xbf - A^\dagger \textbf{b}
\end{align}
and therefore $\xbf \bigtriangledown^\dagger f(\xbf) = \xbf \xbf^\dagger (\Ibb_n + A^\dagger A ) - \xbf \textbf{b}^\dagger A$. The oracle access to entries of $A$ can be used to construct the block encoding of $\varpropto A$, based on the result of \cite{gilyen2019quantum}. The unitary that generates $\textbf{b}$ can be used to construct the block encoding of $\textbf{b}\textbf{b}^\dagger$. Then by the virtue of Lemma \ref{lemma: sumencoding} and \ref{lemma: product}, the block encoding of $\xbf \xbf^\dagger, \varpropto  \Ibb_n + A^\dagger A , \varpropto \xbf \textbf{b}^\dagger  $, and thus eventually can be all combined to yield the block encoding of  $ \xbf \bigtriangledown^\dagger f(\xbf), \bigtriangledown f(\xbf) \xbf^\dagger , \bigtriangledown f(\xbf) \bigtriangledown^\dagger f(\xbf) $. Another application of Lemma \ref{lemma: sumencoding} returns the block encoding of $ \varpropto \xbf \xbf^\dagger  - \eta \xbf \bigtriangledown^\dagger f(\xbf) - \eta \bigtriangledown f(\xbf)  \xbf^\dagger + \eta^2 \bigtriangledown f(\xbf) \bigtriangledown^\dagger f(\xbf)$, which completes an update step. Then the whole process is repeated again, to update another time, and continue until $T$ total iterations, we then obtain the block encoding of $\xbf_T \xbf_T^\dagger$. Using this unitary and apply it to a random state $\ket{\phi}$, according to definition \ref{def: blockencode}, we obtain the state $\ket{\bf 0} \xbf_T \xbf_T^\dagger \ket{\phi}     + \ket{\rm Garbage} $. Measuring the ancilla and post-select on $\ket{\bf 0}$, we obtain the state $\ket{\xbf_T}$, which is a quantum state corresponding to the point of minima of $f(\xbf)$. According to the analysis in \cite{nghiem2025new}, by choosing $T = \log \frac{1}{\epsilon}$, it is guaranteed that $\ket{\xbf_T}$ is $\epsilon$-close to the true minima of $f(\xbf)$, which is also the solution to the linear system. The complexity of this algorithm is $\mathcal{O}\Big( s^2 \frac{1}{\epsilon} \log n  \Big)$.

\section{Review of Method in Ref.~\cite{nghiem2024improved}}
\label{sec: reviewpowermethod}
We review main steps of the improved power method introduced in \cite{nghiem2024improved}, which underlies the lemma \ref{lemma: largestsmallest}. Let $U_A$ denote the unitary block encoding of $A$. Then using Lemma \ref{lemma: product} $k$ times, we can construct the block encoding of $A^k$. Let $\ket{x_0}$ denote some initial state, generated by some known circuit $U_0$ (assuming to have $\mathcal{O}(1)$ depth). Defined $x_k = A^k \ket{x_0}$ and the normalized state $\ket{x_k} = \frac{x_k}{||x_k||}$. According to Definition \ref{def: blockencode}, if we use the block encoding of $A^k$ to apply it to $\ket{x_0}$, we obtain the state:
\begin{align}
   \ket{\phi_1} =  \ket{\bf 0} A^k \ket{x_0} + \ket{\rm Garbage}
\end{align}
Lemma \ref{lemma: improveddme} allows us to construct the block encoding of $\ket{\phi_1}\bra{\phi_1}$, which is:
\begin{align}
    \ket{\phi_1}\bra{\phi_1} = \ket{\bf 0}\bra{\bf 0}\otimes x_k x_k^\dagger + (...)
    \label{d2}
\end{align}
where $(...)$ refers to the irrelevant terms. The above operator is exactly the block encoding of $x_k x_k^\dagger$, according to the definition \ref{def: blockencode}. Recall that we are given $U_0$ that generates the state $\ket{x_0}$, Lemma \ref{lemma: improveddme} allows us to block-encode the operator $\ket{x_0}\bra{x_0}$. We then use Lemma \ref{lemma: product} to construct the block encoding of $ x_k x_k^\dagger \cdot \ket{x_0}\bra{x_0} \equiv ||x_k||^2 \braket{x_k,x_0} \ket{x_k}\bra{x_0}$. The following two results are from \cite{gilyen2019quantum}: 

\begin{lemma}[Corollary 64 of \cite{gilyen2019quantum}  ]
\label{lemma: exponential}
   Let $\beta \in \mathbb{R}_+$ and $\epsilon \in (0,1/2]$. There exists an efficiently constructible polynomial $P \in \mathbb{R}[x]$ such that 
   $$ \Big|\!\Big| e^{ -\beta ( 1-x ) } - P(x)  \Big|\!\Big|_{x\in[-1,1]} \leq \epsilon. $$
   Moreover, the degree of $P$ is $\mathcal{O}\Big( \sqrt{\max[\beta, \log(\frac{1}{\epsilon})] \log(\frac{1}{\epsilon}}) \Big).$
\end{lemma}

\begin{lemma}\label{lemma: qsvt}[\cite{gilyen2019quantum} Theorem 56]
\label{lemma: theorem56}  
Suppose that $U$ is an
$(\alpha, a, \epsilon)$-encoding of a Hermitian matrix $A$. (See Definition 43 of~\cite{gilyen2019quantum} for the definition.)
If $P \in \mathbb{R}[x]$ is a degree-$d$ polynomial satisfying that
\begin{itemize}
\item for all $x \in [-1,1]$: $|P(x)| \leq \frac{1}{2}$,
\end{itemize}
then, there is a quantum circuit $\tilde{U}$, which is an $(1,a+2,4d \sqrt{\frac{\epsilon}{\alpha}})$-encoding of $P(A/\alpha)$ and
consists of $d$ applications of $U$ and $U^\dagger$ gates, a single application of controlled-$U$ and $\mathcal{O}((a+1)d)$
other one- and two-qubit gates.
\end{lemma}
Define $\gamma =  ||x_k||^2 \braket{x_k,x_0}$ for simplicity. We remark that even though the above lemma requires $A$ to be Hermitian, however, for non-Hermitian $A$, it still works on the singular values of $A$ instead of eigenvalues (see Theorem 17 and Corollary 18 of \cite{gilyen2019quantum}). We use the above lemmas to perform the following transformation on the block-encoded operator: 
\begin{align}
    \gamma \ket{x_k}\bra{x_0} \longrightarrow e^{-\beta(1-\gamma)} \ket{x_k}\bra{x_0}+ \sum_{m} P(0) \ket{u_m}\bra{v_m} 
\end{align}
where $\{\ket{u_m},\ket{v_m} \}$ denotes the singular vectors corresponding to zero singular values of $\ket{x_k}\bra{x_0}$. Now we take the above block encoding and apply it to $\ket{x_0}$, and according to definition \ref{def: blockencode}, we obtain the following state:
\begin{align}
    \ket{\bf 0} \Big( e^{-\beta(1-\gamma)} \ket{x_k}\bra{x_0}+ \sum_{m} P(0) \ket{u_m}\bra{v_m}  \Big) \ket{x_0}  + \ket{\rm Garbage} = \ket{\bf 0}  e^{-\beta(1-\gamma)} \ket{x_k} + \ket{\rm Garbage}
    \label{eqn: d4}
\end{align}
where we have used the orthogonality of $\ket{x_0}$ and $ \{\ket{v_m} \} $. Measuring the first register and post-select on $\ket{\bf 0}$, yields the state $\ket{x_k}$ on the remaining register. The success probability of this measurement is $e^{-2\beta(1-\gamma)}$, and as pointed out in \cite{nghiem2024improved}, by choosing $\beta$ sufficiently small, this success probability is lower bounded by some constant, e.g., $1/2$. From $\ket{x_k}$, we use the block encoding of $A$ to apply and obtain the state:
\begin{align}
    \ket{\bf 0} A \ket{x_k} + \ket{\rm Garbage}
\end{align}
Taking another copy of $\ket{x_k}$ and append another ancilla $\ket{\bf 0}$, we then observe that the overlaps: 
\begin{align}
    \bra{\bf 0}\bra{x_k} \big( \ket{\bf 0} A \ket{x_k} + \ket{\rm Garbage}\big) = \bra{x_k}A \ket{x_k}
\end{align}
which is an approximation to the largest eigenvalue of $A$. In order to achieve an additive error $\epsilon$, i.e., 
\begin{align}
    |\bra{x_k}A\ket{x_k} - A_1| \leq \epsilon \\
    || \ket{x_k} - \ket{A_1} || \leq \epsilon
\end{align}
according to \cite{friedman1998error, golub2013matrix}, the value of $k$ needs to be of order $\mathcal{O}\big( \frac{1}{\Delta} (\log \frac{n}{\epsilon} \big)$. In the above procedure, we use the block encoding of $A$ $k$ times, and then use Lemma \ref{lemma: theorem56} to transform to a polynomial of degree $\mathcal{O}( \log \frac{1}{\epsilon} )$ (per Lemma \ref{lemma: exponential}). The overlaps above can be estimated via Hadamard test or SWAP test, incurring a further $\frac{1}{\epsilon}$ complexity for an estimation of precision $\epsilon$. So the total complexity for estimating largest eigenvalue $A_1$, up to $\epsilon$ error is 
$$\mathcal{O}\Big( \frac{1}{\Delta \epsilon} T_A \big(\log \frac{n}{\epsilon}\big) \log\frac{1}{\epsilon}\Big) $$
and the complexity for obtaining $\ket{x_k}$, which is an approximation to $\ket{A_1}$ is $ \mathcal{O}\Big( \frac{1}{\Delta} T_A \big(\log \frac{n}{\epsilon}\big) \log \frac{1}{\epsilon} \Big)$. The above summary completes the details for Lemma \ref{lemma: largestsmallest}, which we left in the main text. \\

In the following, we show how to obtain the operator $A_1 \ket{A_1}\bra{A_1}$ in the Lemma \ref{lemma: extensionlemmalargestsmallest}. Recall from Eqn.~\ref{eqn: d4} above that we obtained the state: 
\begin{align}
    \ket{\bf 0}  e^{-\beta(1-\gamma)} \ket{x_k} + \ket{\rm Garbage} \equiv \ket{\phi}
\end{align}
Lemma \ref{lemma: improveddme} allows us to construct the block encoding of $\ket{\phi}\bra{\phi}$, which is: 
\begin{align}
    \ket{\bf 0}\bra{\bf 0} \otimes e^{-2\beta(1-\gamma)} \ket{x_k}\bra{x_k} + (...)
\end{align}
where $(...)$ denotes irrelevant term. According to Definition \ref{def: blockencode}, the above operator is the block encoding of $e^{-2\beta(1-\gamma)} \ket{x_k}\bra{x_k} $. Now we analyze the term $e^{-2\beta(1-\gamma)} $ and show that for a sufficiently small $\beta$, we have $1- e^{-2\beta(1-\gamma)} \leq \epsilon$. Recall that we defined $\gamma =  ||x_k||^2 \braket{x_k,x_0}$, so apparently $-1 \leq \gamma \leq 1$, which implies $ 1- \gamma \geq 1$. We have that:
\begin{align}
    1- e^{-2\beta(1-\gamma)} &\leq \epsilon \\
    \longrightarrow 1- \epsilon &\leq e^{-2\beta(1-\gamma)}  \\
    \longrightarrow \log (1-\epsilon) &\leq  -2\beta (1-\gamma) \\
    \longrightarrow \log \frac{1}{1-\epsilon} &\geq 2\beta (1-\gamma)
\end{align}
which indicates that $\beta \leq \frac{1}{2(1-\gamma)}\log \frac{1}{1-\epsilon} $. So by choosing a sufficiently small value of $\beta$, we have that $e^{-2\beta(1-\gamma)} \leq 1-\epsilon $, thus implying:
\begin{align}
    || \ket{x_k}\bra{x_k} - e^{-2\beta(1-\gamma)} \ket{x_k}\bra{x_k} || \leq |1-  e^{-2\beta(1-\gamma)} | \leq \epsilon
\end{align}
So the block-encoded operator $e^{-2\beta(1-\gamma)} \ket{x_k}\bra{x_k}  $ is $\epsilon$-approximated to $\ket{x_k}\bra{x_k}$, which is again an $\epsilon$-approximation of $\ket{A_1}\bra{A_1}$ provided $k$ is chosen properly, as mentioned in the previous paragraph. By additivity, $ e^{-2\beta(1-\gamma)} \ket{x_k}\bra{x_k} $ is $2\epsilon$-approximation to $ \ket{A_1}\bra{A_1}$. From the block encoding of $e^{-2\beta(1-\gamma)} \ket{x_k}\bra{x_k} $, we can use Lemma \ref{lemma: product} to construct the block encoding of $ A e^{-2\beta(1-\gamma)} \ket{x_k}\bra{x_k} \approx A \ket{A_1}\bra{A_1} = A_1 \ket{A_1}\bra{A_1} $, thus completing the Lemma \ref{lemma: extensionlemmalargestsmallest}. 

\section{Proof of Lemma \ref{lemma: extensiongradientdescent}   }
\label{sec: extensiongradientdescent}
We remind the reader that the goal is to obtain the block encoding of $\lambda_1 \ket{\lambda_1}\bra{\lambda_1}$, and we have the block encoding of $\xbf_T \xbf_T^\dagger$, which is equivalent to $||\xbf_T||^2 \ket{\xbf_T}\bra{\xbf_T}$.  This block-encoded operator is essentially similar to what we had in Eqn.~\ref{d2}  , therefore, we can follow exactly the same procedure as in previous section (everything below Eqn.~\ref{d2}), and end up obtaining an $\epsilon$-approximated block encoding of $\mathcal{C} \ket{\xbf_T}\bra{\xbf_T}$. As worked out in the main text, by choosing $T = \mathcal{O}(\log\frac{1}{\epsilon})$, it is guaranteed that $|| \ket{\xbf_T} - \ket{\lambda_1}||\leq \epsilon$. Therefore, by additivity of error, we have that  $|| \mathcal{C} \ket{\xbf_T}\bra{\xbf_T}  - \lambda_1 \ket{\lambda_1}\bra{\lambda_1} || \leq 2\epsilon$.

\end{document}